\documentclass[oldversion]{aa}

\usepackage{graphicx}
\usepackage{txfonts}
\usepackage{amssymb}
\usepackage{longtable}
\usepackage{natbib}
\bibpunct{(}{)}{;}{a}{}{,} 

\begin{document}

\titlerunning{Infrared study of Hickson compact groups}
\authorrunning{Bitsakis et al.}

\title{A mid-IR study of Hickson compact groups I :\\
	 Probing the effects of environment in galaxy interactions}

	\author{T. Bitsakis 
		\inst{1}
       \and
     	V. Charmandaris
		\inst{1,2,3}		
	  \and
		  E. Le Floc'h
		\inst{4}
       \and
     	T. D\'iaz-Santos
		\inst{1,2}		
       \and
     	S. K.  Slater
		\inst{5}		
       \and
     	E. Xilouris
		\inst{6}		
       \and
     	M. P. Haynes
		\inst{7}		
}

\institute{Department of Physics, University of Crete, GR-71003, Heraklion, Greece
	  \and
	IESL/Foundation for Research and Technology-Hellas, GR-71110, Heraklion, Greece
	\and
	Chercheur Associ\'e, Observatoire de Paris, F-75014,  Paris, France
	\and
	Laboratoire AIM, CEA/DSM - CNRS - Universit\'e Paris Diderot, DAPNIA/Service d'Astrophysique,  F-91191, Gif-sur-Yvette Cedex, France
	\and
	Department of Physics, Harvard-Smithsonian Center for Astrophysics, Harvard University, Cambridge, MA 02138, USA
	\and
	Institute of Astronomy \& Astrophysics, National Observatory of Athens, GR-15236 Athens, Greece
	\and
	Astronomy Department, Cornell University, Ithaca, NY 14853, USA}

\offprints{T. Bitsakis,  e-mail: bitsakis@physics.uoc.gr}


\abstract{ Hickson compact groups (HCGs) are among the densest galaxy environments of the local universe. To examine the effects of the environment on the infrared properties of these systems, we present an analysis of {\em Spitzer} and{ \em ISO } mid-infrared imaging, as well as of deep ground-based near-infrared imaging of 14 HCGs containing a total of 69 galaxies. Based on mid-infrared color diagnostics we identify the galaxies that appear to host an active nucleus, while using a suite of templates, and fit the complete infrared spectral energy distribution  for each group member. We compare our estimates of galaxy mass, star formation rate, total infrared luminosities, and specific star formation rates (sSFR) for our HCG sample to samples of isolated galaxies and interacting pairs and find that overall there is no discernible difference among them. However, HCGs that can be considered as dynamically ``old'' host late-type galaxies with a slightly lower sSFR than the one found in dynamically ``young'' groups. This could be attributed to multiple past interactions among the galaxies in old groups, that have led to the build up of their stellar mass. It is also consistent with our prediction of the presence of diffuse cold dust in the intergalactic medium of 9 of the dynamically ``old'' groups.}


\maketitle

\section{Introduction}

Compact groups of galaxies were identified as systems of several
galaxies that, owing to their small projected separations and signs of
tidal distortion, appear to be real physical entities of
gravitationally interacting systems \citep{Hickson97}. Since the
discovery in 1877 of Stephan's Quintet, the prototypical group, many 
others have been found by both visual and automated
searches of the Palomar Observatory Sky Survey plates. Although the
criteria for assigning group membership based only on imaging data have 
often been debated, it is now generally accepted that the most complete and
better studied of these samples is the one compiled by \citet{Hickson82}.
His catalog of the so--called Hickson compact groups (hereafter HCGs)
consists of 451 galaxies contained in 100 groups of four or more galaxies, that 
occupy compact configurations within relatively isolated regions
where no excess of other surrounding galaxies can be seen
\citep[see][ for details on the criteria used]{Hickson82}. More 
recently, it has been revealed that the HCG catalog  is incomplete because several compact groups 
are apparent associations of galaxies along the line of sight. 
However, compact group catalogs, produced with redshift information, to select
galaxies that are physically close, are dense \citep{Mamon09}.

Over the past 20 years, various detailed analyses of HCGs have
been performed using multi--wavelength imaging and spectroscopy.
Based on optical morphology alone, it is obvious that the overwhelming
majority of HCGs display an excess of elliptical galaxies ($\sim$31\%
of the total) while their spiral fraction is just 43\%, nearly a
factor of two less than what is observed in the field
\citep{Hickson82}. \citet{Rubin91} have found that two thirds of HCG
spirals display peculiar rotation curves, while \citet{Zepf93b}
report that ellipticals in compact groups are likely to have
irregular isophotes and exhibit lower internal velocity dispersions
for their luminosities, so that they do not lie on the fundamental
plane of isolated field ellipticals. Interestingly, the luminosity
function derived for HCG members is deficient at the faint end
compared to other samples. All these clues are consistent with an
evolutionary pattern of tidal encounters and the accretion of small
companions by the group members. Furthermore, \citet{Mendes94} also
show that 43\% of all HCG galaxies display morphological features of
interactions and mergers, such as bridges, tails, and other distortions.

The HCGs appear to occupy a unique position in the range of galaxy
environments found in the local universe. While their density
enhancements are high, close to those seen in rich clusters, the
overdensities appear to be more locally contained, with a much smaller
population involved in the enhancement. Furthermore, the HCG velocity
dispersions are $ \sigma_V\sim$250 \,~km\,s$^{-1}$, lower than what is
seen in rich clusters but higher than that of typical loose groups. In
addition, a clear correlation exists between the group velocity
dispersion and the elliptical galaxy fraction, with the highest values
of $\sigma_V$ to be found in the most elliptical rich groups
\citep{Hickson88}. 

Despite this progress, several open questions remain.
Are compact groups a transient phenomenon fading
out after the merging of all their galaxies into a giant field
elliptical?   Were they more numerous in the early universe, and can
they account for all, or most, present-day giant ellipticals? Are
they dynamically closed systems, or can they replenish the
intergalactic medium with reprocessed material in the form of diffuse
tails and tidal dwarf galaxies?

A necessary step in determining of the evolutionary state of
HCGs is the complete census and analysis of the member galaxies'
stellar population, gas content, and star formation properties.  Even
though only a few HCGs have been mapped in detail in
neutral hydrogen, single-dish measurements reveal that they are
generally deficient in HI with a median mass of M$_{\rm
HI}\sim2.2\times10^{10}\,{\rm M}_{\odot}$, two times less than what
is observed in loose groups \citep{Williams87,Verdes01}. However, the
molecular gas content of their individual galaxies, as traced by the
CO emission, is similar to that of interacting pairs and starburst
galaxies \citep{Leon98}.  Nearly 40\% of the HCG members for which
nuclear spectroscopy has been obtained display evidence of active
galactic nuclei (AGN), and it has been argued that those contain
copious amounts of dust \citep{Shimada00}. Up to now, however,
there is scant information on the far-IR colors and luminosities of
the individual galaxies of the groups, because the available IRAS
measurements have typically not resolved the groups, especially at 60
and 100\,$\mu$m, providing only a single value and color for the
ensemble of their galaxies.  Despite this limitation, it has been shown
that compact groups have warm far-IR colors, similar to those of
merging gas-rich galaxies \citep{Zepf93a}.  If one recalls that HCGs
are in fact deficient in spirals and that the far-IR flux of field
ellipticals is typically below the IRAS detection limit, one may
speculate that vigorous star formation activity may actually take
place in some locations within the groups. Could this activity be in
circumnuclear regions of weak, enshrouded AGN because of the dynamical
torques exerted in the gas from small nuclear bars funneling it to the
center \citep{Combes01}? Would it be possible instead that there is
star formation due to shocked compressed gas outside the main bodies
of galaxies as seen in Stephan's Quintet \citep{Xu99, Appleton06}?

A first analysis of the mid-infared properties of 12 HCG using Spitzer imagery 
has been presented by \citet{Johnson07} and \cite{Gallagher08}. These authors use 
near- and mid-IR color diagnostics to reveal a possible  excess thermal emission due 
to an active nucleus (AGN). They also separate the groups into three types based on 
the ratio of their neutral hydrogen to dynamical mass, and find that most groups 
are either gas-rich or gas-poor, while very few groups are found in an intermediate state. 
As expected, gas-rich groups are the ones that 
are more actively forming stars. In addition, an absence of intermediate stage 
groups is apparent, which would be consistent with a rapid evolution of the 
group members, probably the result of dynamical interactions.

In this paper we present the first results of our detailed analysis of the deep 
near-IR, mid-IR, and far-IR  imagery of 14 HCGs and place them in the context of 
similar results for samples of isolated galaxies and early-stage interacting 
systems. In Section 2, we describe our data and reduction processes, along 
with the control samples used. In Section 3, we describe how we model the infrared 
spectral energy distribution (SED) of all galaxies in the groups. Our results 
on the physical properties of each galaxy 
such as star formation rate (SFR), stellar mass, dust content, and specific 
star formation rate (sSFR) are presented in Section 4. We discuss the implications 
of these findings in Section 5, and present our conclusions in Section 6.

\section{Observations and data reduction}

Our sample of 14 HCGs  contains 69 galaxies, 31 of which are early-type and 38 
late-type galaxies. It was selected in a random fashion from the original Hickson catalog (see Table 1 of Hickson 1982)
and is presented in Table 1. The only constraint applied was 
that the group members had to be separated enough to allow more accurate photometry, but also contained enough within 
$\sim5' \times5'$ to be efficiently imaged by the Wide-field Infrared Camera 
(WIRC) at the Palomar 5m telescope. This constrains our sample to nearby groups typically with 
5 or 6 members. Despite the small numbers, the sample can be considered representative 
of Hickson's catalog having an early-type galaxy fraction of $\sim40\%$ and a 
spiral and irregular galaxy fraction  of just $\sim60\%$ of the total. Furthermore, as we discuss in section 4.2, Kolmogorov-Smirnov (KS) tests indicate  that the mass distribution of our subsample, estimated by their Ks-band luminosites, is representative of the whole HCG  sample. Nine galaxies of our sample are classified as AGN based on optical spectroscopy, even though this number  is probalby a lower limit given the sparse spectroscopic coverage. In Figures 1 and 2, we show typical images of one group of our sample, HCG 57, using the J, H, Ks, and IRAC filters.

\begin{figure}
\includegraphics[scale=0.3]{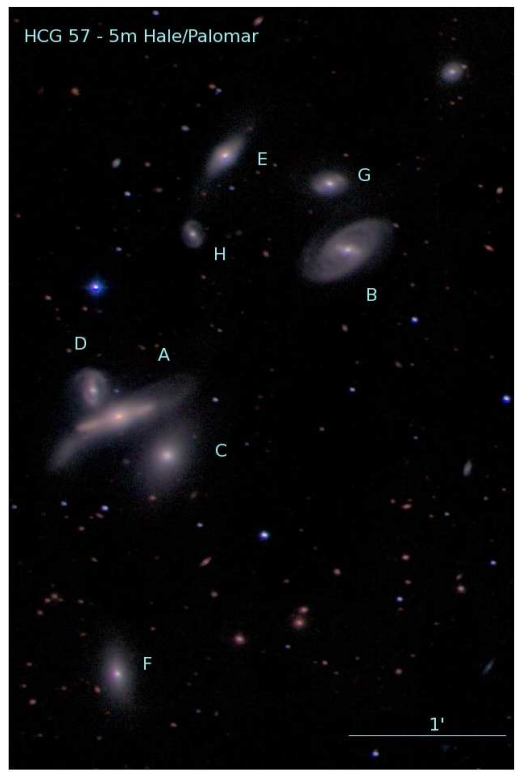}
\caption{A ``true color'' image of HCG57 based on the Palomar  J,H, and Ks data \citep{Slater04}. This is the largest group of our sample containing 4 early and 4 late-type galaxies. The galaxy triplet consisting of members A, C and D is also known as Arp 320. Galaxy A is classified as an AGN.} 
\end{figure}

\begin{figure}
\includegraphics[scale=0.4]{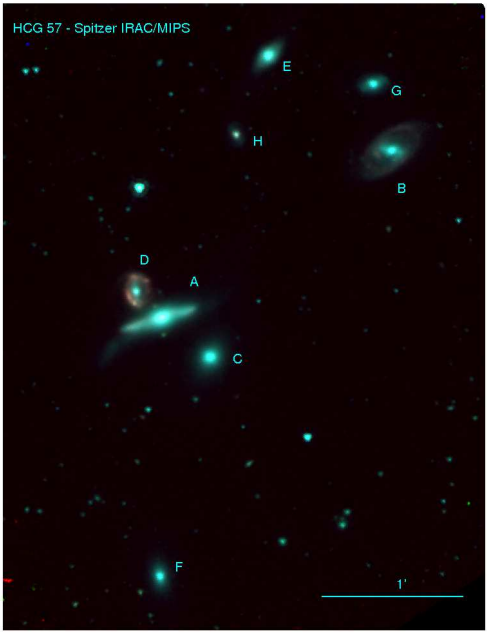}
\caption{A Spitzer/IRAC ``true color'' image of the same group. The blue channel traces the $3.6\mu$m emission, the green the $4.5\mu$m, and red the  $8.0\mu$m. Galaxies A, B, and D have very luminous spiral arms, and when examined more closely, they appear to contain some bright mid-IR spots.}
\end{figure}

\subsection{Near-infared observations}

Deep near-IR observations were obtained with the Wide Field Infrared Camera (WIRC) of the 5m 
Hale telescope at Palomar, during several observing runs in the 2002 and 2003 observing periods. 
All groups were imaged in the J, H, and Ks bands for an on-source time of 20 minutes per 
filter \citep{Slater04}. The 20 minute exposures were taken in multiple shorter segments 
to avoid saturation by the sky brightness.  The telescope was dithered in a 3- or 
5-position pattern during the exposure. Due to sizable telescope nodding as well as dithering, 
it was necessary to perform astrometry  on each of the shorter exposure images separately.  
This was done with the software-package WCSTools \citep{Mink02}, matching foreground stars 
to the USNO A2.0 and 2MASS point source catalogs.  Once a World Coordinate System (WCS) 
was established for each image, 
they were pixel-resampled and coadded using the Swarp package \citep{Bertin08}.  Source 
extraction was performed with SExtractor \citep{Bertin96}. The flux was calibrated to a 
level of 0.1mag using near-IR standards, and it was bootstrapped  to 2MASS foreground stars, 
which were detected in both our images. Our 1$\sigma$ sensitivity limit was $\sim$ 21.5 mag 
arcsec$^{-2}$ in J and H bands and  $\sim$20.5 mag arcsec$^{-2}$. A near-IR
image of the group HCG\,57 is presented in Figure 1.

\subsection{Mid-infrared Spitzer observations}

Observations of these groups were obtained between 2008 January and 2009 March in the $3.6, 
4.5, 5.8$, and $8.0 \mu$m broadband filters of the Spitzer Infrared 
Array Camera \citep[IRAC,][]{Fazio04} and the $24 \mu$m band of the Multiband Imaging 
Photometer for Spitzer \citep[MIPS,][]{Rieke04} (PID: 40459). The IRAC observations were 
obtained with a 270sec exposure per filter for each set of cameras (see Fig. 2). All groups had an 
angular size small enough to fit within the $5' \times 5'$ field of view of IRAC. The 
on-source time for the 24$\mu$m MIPS observations was 375.4sec.

The IRAC postpipeline Basic Calibrated Data were downloaded and mosaicked 
onto the same grid with a pixel scale of $1.2'' \times 1.2''$. The 1$\sigma$ sensitivity 
limits varyied slightly for each group owing to changes in the background. 
Typical $3\sigma$ limits were measured to be $\sim0.04$ MJy sr$^{-1}$, $\sim0.05$ MJy 
sr$^{-1}$, $\sim0.2$ MJy sr$^{-1},$ and $\sim0.2$ MJy sr$^{-1}$,  for the 3.6, 4.5, 5.8, and  
8.0$\mu$m filters, respectively. Since most of the galaxies have disturbed morphologies, simple fixed aperture 
photometry was often not the most appropriate method of measuring the object flux.  Thus, we carefully calculated 
the isophotal contours around each source to properly account for variations in the shape of the emitting 
region. Examining the local background  for each galaxy we defined a limiting isophote 3$\sigma$ above the overall background 
and measured the flux within this region, after  subtracting the corresponding sky.  Even this method was challenging in some cases, so we defined the physical extent of each galaxy by hand using  a polygon,  and performed the photometry accordingly. The fluxes we report are for the same regions for all IRAC, and MIPS, images. Finally, we applied an extended source correction to our photometry by  multiplying our measured fluxes with a correction factor of 0.91, 0.94, 0.71, and 0.74 for the corresponding 1 to 4 IRAC bands as indicated in the IRAC data handbook.

The MIPS  postpipeline images have a pixel scale of $2.45'' \times 2.45''$, 
and their depth was measured as $\sim0.3$ MJy sr$^{-1}$ or $\sim250\mu$Jy for a point source.
In several cases the nuclei of galaxies were saturated and the frames were corrected 
using the IDP3 of IDL\footnote{IDP3 is the Image Display Paradigm 3 package available for the 
Interactive Data Language (IDL)}. We estimate that our final uncertainties in the IRAC and MIPS 
photometry are $\sim5\%$.

To ascertain variations between the nuclear and total mid-IR SED of each galaxy, we also performed
nuclear photometry for our sources using the smallest aperture possible. The motivation behind this work 
was to examine whether the observed SEDs of some sources are dominated by the contribution of a strong nuclear emission, possibly due to the presence of an AGN.  The apertures used were 3 pixels (3.6$"$) in diameter for the IRAC images and 5 pixels (12.2$"$) for 
the MIPS, based on the size of the corresponding point spread function (PSF).  

\subsection{Mid-infrared 12$\mu$m ISO/CAM data}

Archival observations for a fraction of the galaxies were retrieved from the Infrared 
Space Observatory (ISO) data archive.  The observations were performed with the ISO/CAM  instrument \citep{Cesarsky96}, 
using the LW10 filter (Program HCGROU-A). The LW10 filter is centered on 
$\lambda=11.7\mu$m, and covers the $8.5-15.5\mu$m range so that it matches the 
IRAS 12$\mu$m bandpass. The pixel size was $6''\times6''$, and it is larger than the 
PSF of the telescope at this wavelength. The total on-source exposure time for 
the galaxies varied between 200 and 300sec.

The ISO/CAM data were analyzed with the CIR data processing software, an implementation 
of the algorithms of \citet{Starck99} developed by P. Chanial. The standard ISO/CAM data 
reduction procedure was followed. First the data, were read into the CIR environment. Then 
the ``correct-dark-vilspa" routine was applied to correct for the dark current. This was 
followed by applying the multiresolution algorithm  ``correct-glitch-mr". Next, the 
``correct-transient-fs" routine, based on the Fouks-Schubert transient correction method 
\citep{Coulais00}, was applied to the data to remove transient effects. The frames were 
flat-fielded using the CIR's library flat field, and subsequently combined to produce a 
final image. Some masking on the edges of the images was performed to correct for bad 
pixels. Photometry was performed by selecting the regions of galaxies with emission 
greater than 3$\sigma$ above the sky background noise. We estimate that the absolute 
uncertainty of our photometric measurements is $\sim$\,10$\%$, mainly owing to 
errors in the correction of the detector transient effect.

\subsection{Comparison samples}
\begin{figure}
\resizebox{\hsize}{!}{\includegraphics[scale=0.5]{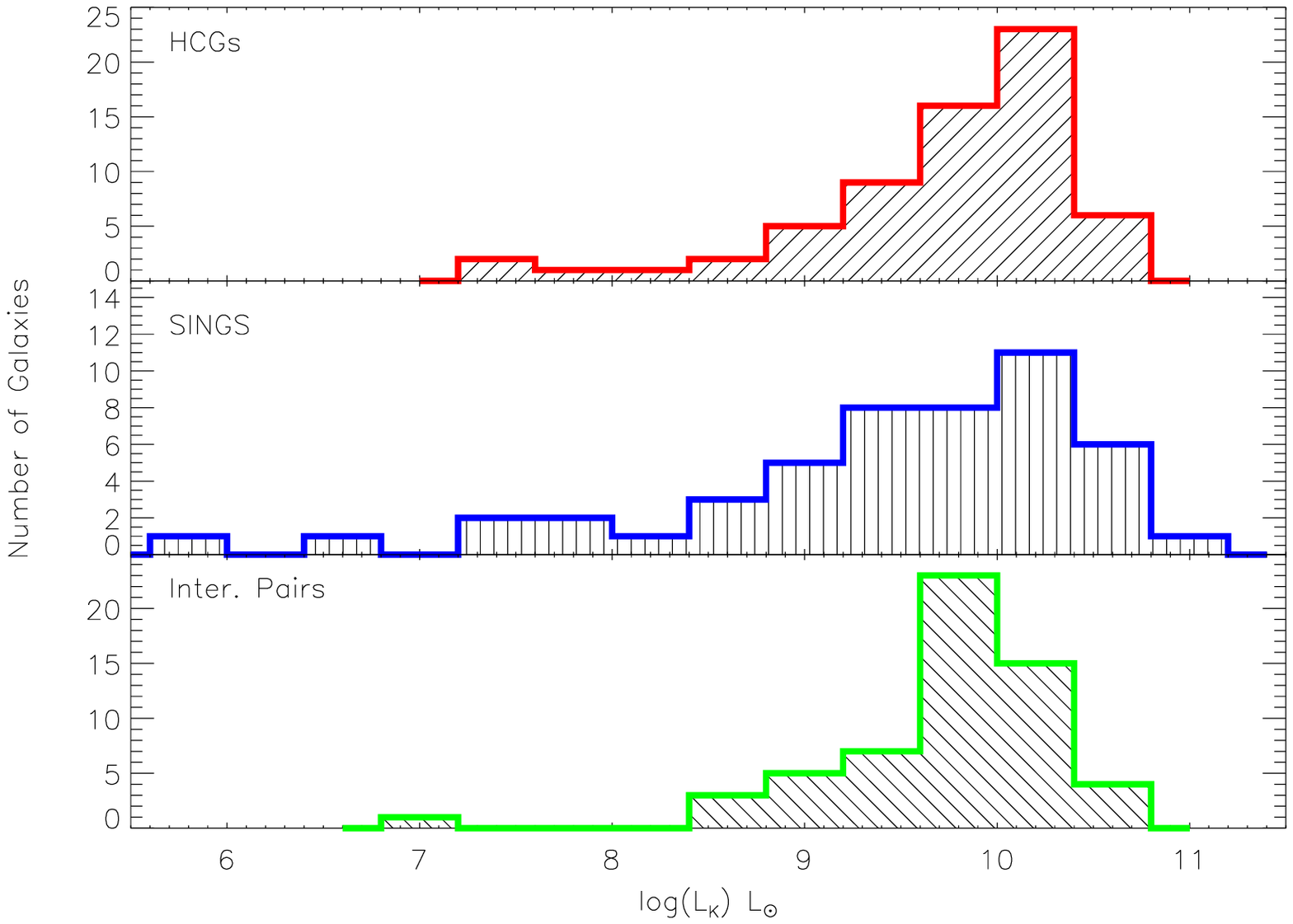}}
\resizebox{\hsize}{!}{\includegraphics[scale=0.5]{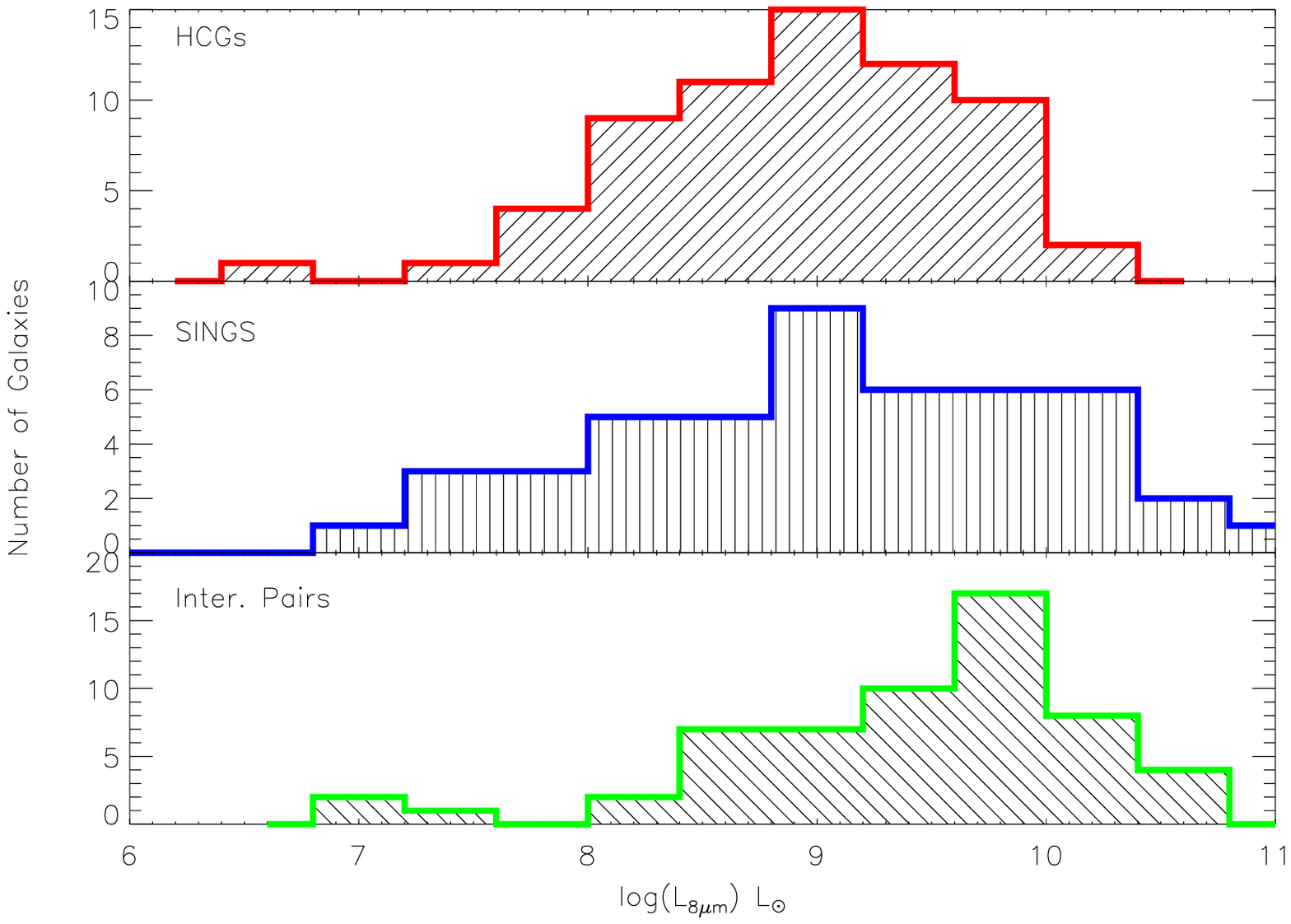}}
\resizebox{\hsize}{!}{\includegraphics[scale=0.5]{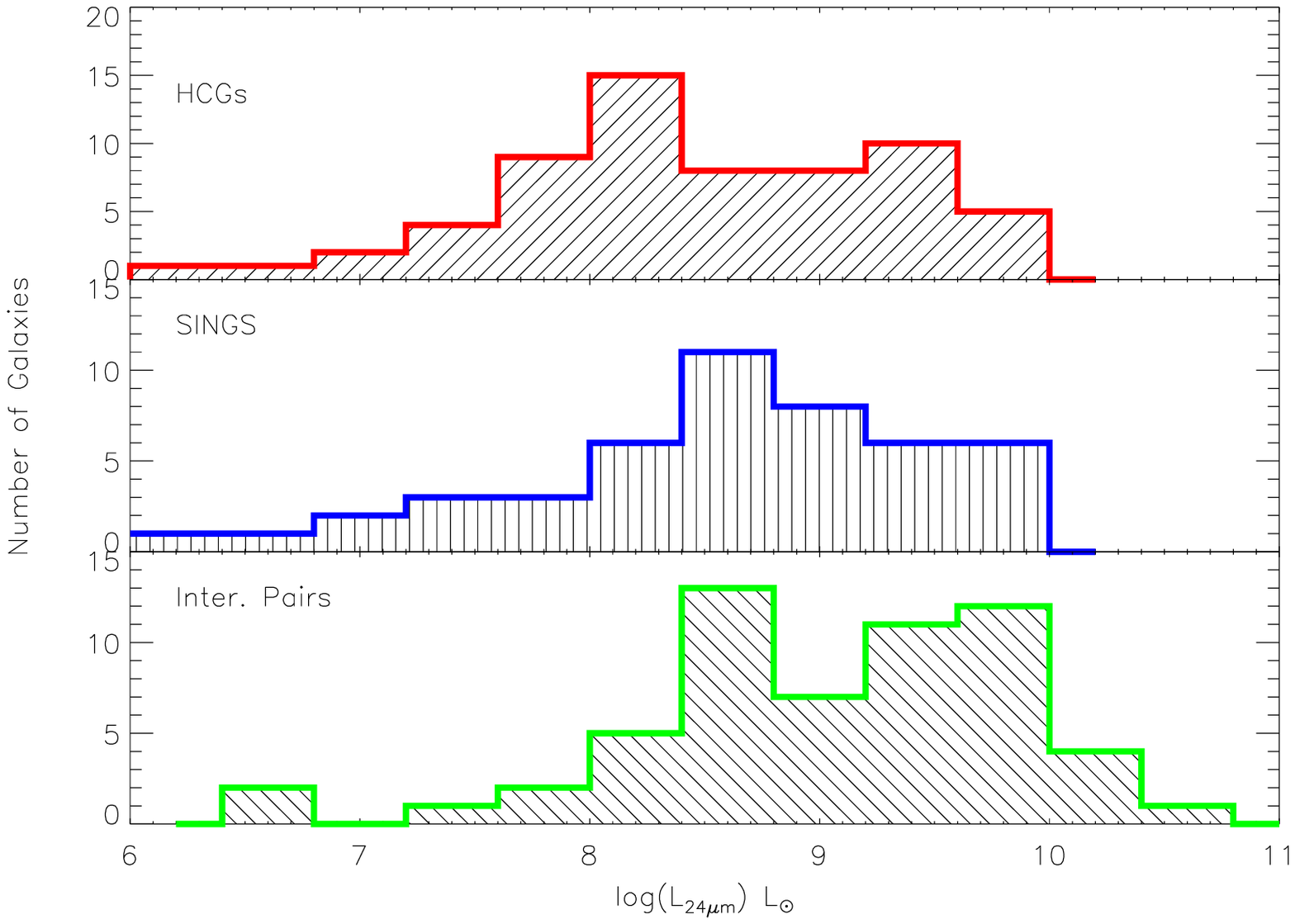}}
\caption{Comparison of the K-band, 8$\mu$m, and 24$\mu$m luminosities 
of our samples: HCGs (top panels), field galaxies (middle panels), and interacting pair galaxies (lower panels).}
\end{figure}

We compare our findings to the published 
mid-IR data on the HCG of the \citet{Johnson07} sample, as well as on two ``control" samples, 
for which a similar type of data exist. One is the nearby ``normal" sample of 75 field galaxies 
from the Spitzer Infrared Nearby Galaxies Survey \citep[SINGS;][]{Kennicutt03, Dale05}. 
The SINGS sample was selected to cover a wide range in Hubble type and luminosity. Most 
objects are late-type systems that have angular sizes between 5$'$ and 15$'$.  It also 
contains four early-type galaxies (NGC 855, 1377, 3773, and 4125) with the last one being 
a LINER. The second comparison sample contains the 36 nearby (v$<$11.000 km s$^{-1}$) early 
stage interacting galaxy pairs of \citet{Smith07a}. The galaxies in the Smith sample are 
tidally disturbed and fairly extended, with linear sizes $>3'$. We note that 13 
galaxies of this sample are classified as LINERs and 3 as Seyferts. The motivation behind the 
comparison of our HCG sample to these samples was to investigate how the group environment affects 
the infrared  properties of galaxies, in contrast to isolated systems, because of a single companion galaxy. The basic properties of the 
three samples are presented in Figure 3. The mid-IR activity in groups, 
as traced by the 24$\mu$m and 8$\mu$m luminosity, is similar in all three samples, suggestive that as 
samples  there is no substantial difference in their SFR. This has already been pointed out by \citet{Smith07a}, who 
find that there was no more than a factor of two difference between the SFR in isolated (SINGS) galaxies 
and early stage interacting pairs. The K-band luminosities, hence the stellar mass distributions, are also similar with the exception of  the interacting pairs, who do not extend to masses below $\sim10^{8}$M$_{\odot}$.

\section{Spectral energy distribution fitting}

To estimate the physical properties of the galaxies in our sample in a systematic way, we first  fit the observed SED of each galaxy with a set of theoretical and empirical SED models from {\it Le Phare}\footnote{For more information on {\it Le Phare} visit the web page of the package at http://www.cfht.hawaii.edu/$\sim$arnouts/LEPHARE/cfht$\_$lephare/lephare.html} \citep[see][]{Arnouts07,Ilbert09}. We compared the observed near-IR J, H, and Ks fluxes, as well as the mid-IR fluxes from Spitzer and ISO to the ones predicted using the {\it Le Phare} template SEDs.

Using multi-wavelength template libraries, 
{\it Le Phare} is able to perform a two-component galaxy SED modeling by decomposing the fit of each source into the contribution of its stellar populations and the contribution from thermal dust emission. The near-IR observations is strongly dominated by stellar blackbody emission, with a possibly small contribution from very hot dust in the case of an active nucleus (AGN). In late-type, star-forming galaxies, two of the IRAC bands, centered on 5.8 and 8$\mu$m, may contain the contribution of polycyclic aromatic hydrocarbons (PAHs), and finally the $24\mu$m luminosity traces thermal emission from dust mostly heated in embedded star-forming regions. Consequently the templates used to fit the data should be compatible with those characteristics. A first fit of the near-IR J, H, and Ks photometry was obtained using pure stellar templates taken from the library of SEDs used by \citet{Ilbert09} to derive galaxy photometric redshifts. The goal of this initial step was to estimate the contribution of the underlying stellar component to the emission observed in the Spitzer bands at longer wavelengths. This contribution was subsequently subtracted from the IRAC and MIPS fluxes assuming an extrapolation of a typical stellar blackbody  emission, and the resulting fluxes were fitted between 5 and 24$\mu$m using a separate library of IR SEDs. For the IR SEDs, we used  a modified version of the IR template library of \citet{Chary01}, which includes 105 star-forming galaxy IR SEDs over  a broad range of total IR luminosities. These templates cover the optical to radio wavelength range, so  their stellar component was systematically removed prior to the fitting with  {\it Le Phare} to reproduce only the thermal dust component of our sources. For each galaxy of our sample, the best total fit was then obtained by co-adding the fit derived for the stellar component and the one obtained in the mid-IR. Galaxies with very low $24\mu$m emission, mostly early-type systems, were only fitted with a single stellar component, since subtracting it from the 24$\mu$m flux would introduce a large uncertainty to the final estimate of the thermal dust emission.

In Figure A.1 of the appendix we present the infrared SEDs of each galaxy along 
with the best-fit model from {\it Le Phare}. Each panel corresponds  to one group; groups with 
more than six members are plotted in two figures. The galaxies are labeled with their HCG number, 
type, and spectral classification,  if known. The vertical error bars indicate the uncertainty 
of each measurement, while horizontal bars show the bandwidth of the corresponding filter.

Most SED fits are fairly good 
as is evident from the plots and the corresponding reduced $\chi^{2}$ values, which vary between 1 and 11 and are marked in each plot. A number of galaxies display mid-IR fluxes consistent with strong PAH emission and often warm dust, a sign of enhanced star formation activity. There are, though, some galaxies (such as HCG26a, 26e, or 57a) that  are not fitted as  well by the templates. These galaxies display an excess emission in the Ks-band and/or mid-IR, which is suggestive of some type of nuclear activity, possibly from  an AGN. Even if star  formation and thus PAH emission were present in these systems, the nuclear thermal dust emission  would dominate the emitted flux, resulting in rather flat or rising mid-IR spectrum.

We also compared our galaxy classification based on the templates of the SED fit to existing optical morphology classification.
Our data, as do the SED fits, indicate that some of the galaxies, classified as 
early-type based on the optical images, do have excess dust emission. These are 
HCG26b, 37c, 37e, 55a, 55b, 55d, 56b, 56d, 56e, 71b, 79b, 95b, and 95d. This implies that 
either embedded star formation or accretion of gas-rich dwarf galaxies have fueled these 
systems with enough dust and gas  to emit strongly in the mid-IR.  All of them seem to 
have late-type galaxy SEDs in the infrared. Three (HCG37c, 56b, and 56d) are spectroscopically 
classified as AGN in the optical, while others exhibit a mid-IR SED that is suggestive 
of emission from an AGN. Interestingly, there are also galaxies that have IR  SEDs 
consistent with that of an early-type galaxy. These include HCG38d and HCG95b, 
which are nonetheless classified as late-type in the optical. This could be due to the loss of 
a large fraction of their gas and dust, probably because of interactions, which would be the result of the 
absence of far-infrared emission. One of the galaxies in our sample, HCG 40f, which did 
not have a classification in the literature, has been fitted with an E1 galaxy template, consistent with its morphological appearance in the optical and infrared images.

It is known that the Ks band luminosity is a good tracer of the stellar mass because light in this band is dominated by the emission of low-mass stars, which are responsible for the bulk of stellar mass in galaxies. Similarly, the $24\mu$m emission is a good tracer of star formation because it originates in regions of dust mainly heated by young stars.
We used the observed Ks-band and $24\mu$m luminosities (L$_{\rm K}$ and  L$_{24\mu m}$) to estimate the corresponding stellar mass and SFR for each galaxy. To obtain the galaxy mass, we used the \cite{Bell03} relation where  the mass, in  M$_{\odot}$, is (0.95$\pm$0.03)$\times$ L$_{\rm K}$(L$_{\odot}$), where L$_{\rm K}$ is in units of K-band solar luminosity (L$_{\rm K, \odot}$=4.97$\times10^{25}$W), with systematic  errors of $\sim30\%$, owing to uncertainties in the star formation history and dust.  The SFR was estimated from  the $24\mu$m by the calibration developed by \citet{Calzetti07}  for the SINGS sample :\\
SFR(M$_{\odot}$yr$^{-1}$) = 1.27$\times10^{-38}$(L$_{24\mu m}$ (erg sec$^{-1}$))$^{0.885}$. \\
Our mass and SFR estimates are reported in Table 2 along with a measure of the sSFR of our sources  defined as the ratio of SFR over the galaxy stellar mass.

\section{Results}

\subsection{IRAC colors and AGN diagnostics}

\begin{figure}
\resizebox{\hsize}{!}{\includegraphics[scale=0.5]{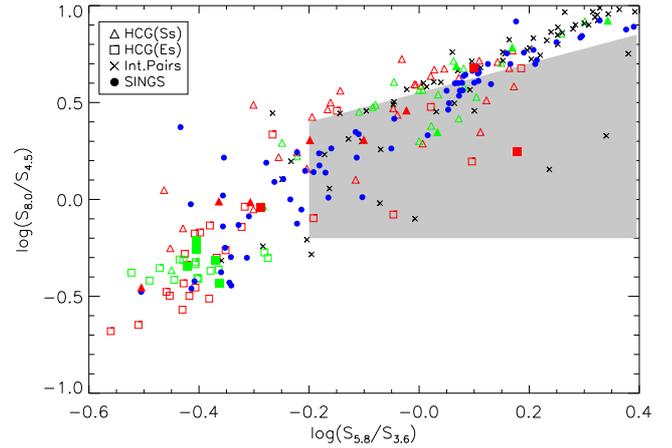}}
\caption{The Lacy et al. (2004) IRAC color diagnostic. Red symbols indicate the galaxies of our HCG sample, green  those of the \citet{Johnson07}  sample, black crosses the interacting pairs of the \citet{Smith07a}, and blue filled circles the SINGS galaxies.  Late-type galaxies (spirals) are marked with triangles, while squares denote early-type galaxies (ellipticals). Filled squares and filled triangles denote that the corresponding galaxy is classified as an AGN based on optical spectroscopy. According to the mid-IR colors, the AGN candidates are those found in the gray shaded area. The gap in the distribution of the galaxies between those on the lower left side of the plot and those on the upper right side, mentioned by  \citet{Johnson07}, may also be present in data, but it is not as clear.}
\label{fig_4}
\end{figure}
\begin{figure}
\resizebox{\hsize}{!}{\includegraphics[scale=0.5]{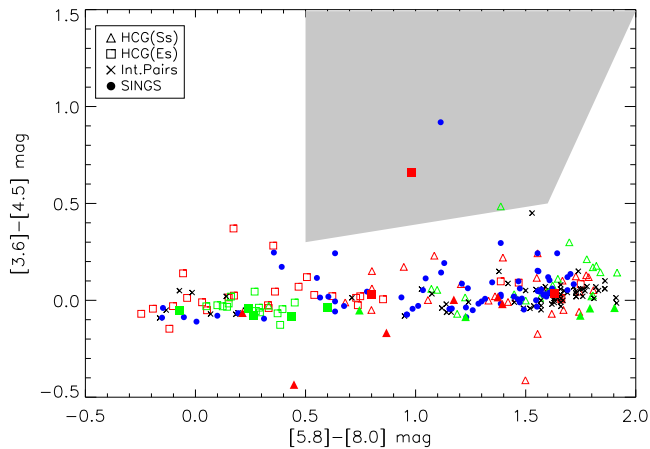}}
\caption{The Stern et al. (2005) IRAC color diagnostic. The symbols are as  in Fig. 3. The gas-poor galaxies are located in the lower left portion and the gas-rich in the lower right. Only one candidate, HCG\,56b (a Seyfert 2), is found within the AGN candidate area.}
\label{fig_5}
\end{figure}

AGN are often deeply enshrouded in dust that may heavily absorb the emitted radiation, in particular in the UV,
optical, and even in the X-rays in some extreme cases. The energy is then re-emitted isotropically at longer wavelengths
in the mid- and far-IR \citep[see][ and references therein]{Charmandaris08, Elitzur08}. As a result, active galaxies often display an 
excess of continuum emission at wavelengths $2\mu m< \lambda <15\mu$m, compared to normal or 
starbursting galaxies. In addition, there is a weak PAH emission in their mid-IR spectrum, which is attributed to the destruction of  their carriers by the strong radiation field surrounding the AGN 
\citep[i.e. see][]{Weedman05, Brandl06, Smith07b}. We can use our Spitzer 
infrared photometric measurements to probe for an AGN among the galaxies 
in our sample. \citet{Lacy04} have defined a locus in  the IRAC  [8.0-4.5]  vs [5.8-3.6] 
color-color plot populated by AGN which are dominant in the infrared. 
Similarly, \citet{Stern05} propose using of  the IRAC  [3.6-4.5]  vs [5.8-8.0] 
colors to identify their AGN candidates. Among our galaxies optical spectra were available for 24
\cite{Shimada00}, with 15 classified as \ion{H}{II} galaxies and 9 as AGNs, 
(3 as Sy2 and 3 as LINERs).

In Figure 4, we present the IRAC  colors for our sample following the \citet{Lacy04} criteria. 
We separate our sample in early and late-type galaxies, and identify the galaxies that we 
already know from optical spectroscopic observations host an active nucleus. 
We also include in Figure 4 the galaxies from the \citet{Johnson07} HCG sample, as well as the SINGS and interacting galaxy samples. By construction, the galaxies located in the 
lower left of the color-color plot are those dominated by the stellar photospheric emission.
As a result, they are expected to have very weak dust or PAH emission, if any. Most of our early-type 
galaxies lie in this region. On the other hand, galaxies found in the upper right quadrant of the 
figure should have strong PAH features, as well as some hot dust contribution, because of intense 
star formation, and/or AGN activity. Furthermore, as we discuss in section 4.4, most of the galaxies in this quadrant are spirals found in spiral-rich groups. According to  \citet{Lacy04}, galaxies located within 
the shaded area are AGN candidates. A total of 16 galaxies in our sample ($\sim23\%$) are AGN candidates 
based on this criterion, almost half of which are found in previous optical spectroscopy studies \citep{Shimada00,Coziol00}. These are 11 late-type systems (HCG 37b, 38b, 38c, 47c, 54abcd, 
57d, 71a, 79a) and 5 early-type systems (HCG 40f, 56b, 56e, 71b, and 95b). Of those galaxies, four are 
spectroscopically classified as AGN, while eight are classified as \ion{H}{II} galaxies. In addition, as we can 
see from their mid-IR SEDs in Figure A.1, several of them do show signs of active nuclei, such as a flat rising continuum. It is possible, though, that  either because of obscuration or because some may harbor low-luminosity AGNs (LLAGNs),  the colors of the AGN are diluted when we perform the photometry over the whole galaxy. To better probe the properties of the nuclear emission, we can examine the nuclear SEDs of our sample, which was obtained using circular apertures with a diameter of 3.6$\arcsec$ in the  IRAC images and 12.2$\arcsec$ in the MIPS 24$\mu$m band. Among the AGN candidates based on the Lacy plot, the ratio of  the nuclear to their total flux is not constant in all mid-IR bands, in five galaxies, HCG 37b, 47c, 56b, 71a, and 79a, but it peaks at  $4.5\mu$m. This suggests that the nuclear contribution which possibly caused by hot dust heated in near sublimation temperatures by an AGN, contributes substantially in this band.  On the other hand, nine galaxies, HCG 38b, 38c, 54abcd, 56e, 71b, and 95b, have a rising nuclear SED and a constant ratio of nuclear to total flux in all IRAC bands. This would be consistent with their being star-forming  \ion{H}{II} galaxies. 

Contrary to SINGS galaxies, which are distributed fairly uniformly along the diagonal  of the color-color plot in Figure 4 and the interacting galaxy pairs which are located in the upper right quadrant due to the fact that they contain only late-type galaxies, there might be a small gap in the distribution of  the HCG galaxies at the position log(S$_{5.8}$/S$_{3.6}$) $\simeq -0.2$ and   log(S$_{8.0}$/S$_{4.5}$)$ \simeq 0$. This gap is identified in Figure 11 of  \citet{Johnson07}, even though it is more obvious there. These authors combine the mid-IR colors with estimates of the neutral hydrogen mass and dynamical mass of the their HCG sample and  suggest that this small gap separates the gas-rich and gas-poor groups. The explanation is consistent with the notion that gas-poor groups will mostly contain early-type galaxies where PAH emission will be suppressed by low star formation activity; and they will populate the lower left part of the plot, while late-type gas-rich galaxies will be found on the upper right. The absence of HCG galaxies with intermediate mid-IR colors would imply that their evolution from the gas-rich to the gas-poor state is more rapid than what is observed in field galaxies, and it could be understood as the result of more frequent dynamical interactions between galaxies in groups. As a result of the interactions, tidal forces drive the gas in their nuclei and create instabilities in their disk, forcing them to use their gas in a more accelerated fashion.

In Figure 5, we also examine how  the \citet{Stern05} color-color diagnostic is applied in our data. As was the case in Figure 4, galaxies located in the lower left region of the plot have SEDs dominated by the stellar photospheric emission, while the galaxies found on the right should display strong PAH and hot dust emission. The shaded area denotes the colors of potential AGN candidates. When we compare this to Figure 4, we notice that only one galaxy, HCG56b, is identified as an AGN candidate. However, as discussed in the  detail by \citet{Donley08} the Stern criteria may suffer in low-z because of contamination from star-forming galaxies. Consequently, in this work we adopt the ones fulfilling the Lacy criteria as mid-IR identified AGN.

\subsection{Star formation activity in Hickson compact groups}

As mentioned in the previous section, one would expect that galaxies in groups would have enhanced SFR, because of tidal interactions. The $24\mu$m emission is a good tracer of the SFR and following  the approach of  \citep{Calzetti07}, we calculated the SFR for each galaxy of our groups in Section 3. In Figure 6a, we plot the sSFR  as a function of  the mass of each galaxy, which was estimated using the observed K-band luminosity  \citep{Bell03}. We include 57 the galaxies of our sample that were detected in 24$\mu$m. In the same figure 61 normal  galaxies  of the SINGS sample are presented, along with 30 of the interacting pairs of \citet{Smith07a} for which integrated K-band 2MASS photometry was  available. For the interacting pairs, the average sSFR of the two components is given. Galaxies found over the dashed line, have SFRs greater than 1M$_{\odot}$yr$^{-1}$. Galaxies in interacting pairs display a median sSFR of  $\sim$2.96$^{+2.70}_{-1.41}$$\cdot10^{-11}$ yr$^{-1}$. This is slightly higher than what is found in HCGs and field galaxies, which have a corresponding value of $\sim$1.68$^{+1.41}_{-1.16}$$\cdot10^{-11}$ yr$^{-1}$. However, given the scatter seen in each population there is no strong statistical difference among them. As  mentioned in Section 2.3, the interacting pair sample does not extend to the lower galaxy masses of the two other samples. However, even when we restrict the galaxy masses to a range that is identical in all three samples, an analysis using Kolmogorov-Smirnov test still indicates that there is no statistical difference between the samples. In Figure 6b, we plot the sSFRs as a function of the total infrared luminosity (L$_{\rm IR}$). For similar infrared luminosities, HCG members seem to have sSFRs similar to field galaxies and slightly lower than interacting pairs.  How is this understood given that HCGs are also interacting systems? 

To better address  this issue, we decided to separate the late-type galaxies of the HCGs into two categories depending on whether the group they belong to has a high or low fraction of late-type galaxies. We classify a group as ``spiral-rich'' if fewer than 25\% of its galaxies are early-type. Conversely, a group is ``spiral-poor'', or elliptically dominated,  if fewer than 75\% of its galaxies are late-type.  The  sSFR is a tracer of the star formation history of a galaxy, and galaxies in compact groups do experience multiple encounters with the various group members. Consequently we would expect that if the group is dynamically ``young'' (that is, its galaxies are gravitationally interacting for the first time) it is more likely  to be dominated by late-type galaxies. Furthermore, these spiral galaxies would have not built up much of their stellar mass, but would have larger amounts of gas and dust, as well as higher SFRs and sSFR. On the other hand, if the group is dominated by ellipticals, it could be dynamically ``old'',  since  interactions and possible merging of its members over its history would have led  to the formation of some of those ellipticals. As a result, the spirals in these groups could have already built some of their stars, and their sSFR would be lower.

\begin{figure}
\resizebox{\hsize}{!}{\includegraphics[scale=0.5]{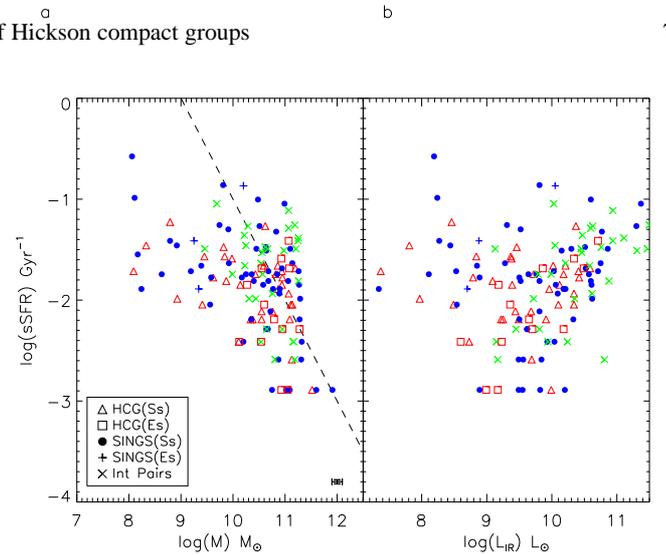}}
\caption{\textbf{a)} Specific star formation rate as a function of stellar mass. The red triangles are HCG late-type galaxies and the red squares are the HCG early-type galaxies detected at 24$\mu$m. Galaxies found in ``spiral-rich" groups are marked with filled symbols. Blue filled circles and  blue crosses indicate SINGS late- and early-type galaxies, respectively. The green x's indicate the interacting galaxy pairs of \citet{Smith07a}. Galaxies to the right of the dashed line have SFR greater than 1M$_{\odot}$yr$^{-1}$.
\textbf{b)} Specific star formation rate as a function of  L$_{\rm IR}$. Symbols as before.}
\label{fig_6}
\end{figure}

\begin{figure}
\resizebox{\hsize}{!}{\includegraphics[scale=0.5]{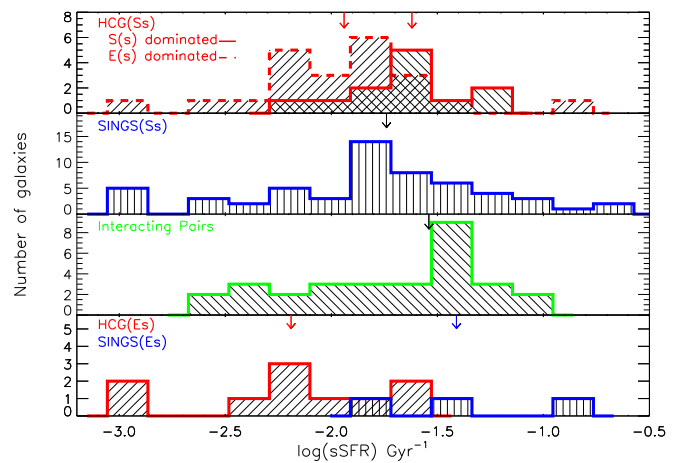}}
\caption{ Histograms of the sSFRs in the samples used. The top plot displays a histogram of the late-type galaxies found in spiral-rich groups marked with the solid line, as well as a histogram of the late-type galaxies in elliptically dominated groups marked with the dashed line. The second is a histogram of the SINGS late-type galaxies, and the third is the corresponding one for the interacting galaxy pairs. Finally, the bottom plot displays  the histogram of the early-type galaxies found in HGCs (in red), as well as the SINGS early-type galaxies (in blue), which are detected at $24\mu$m. The arrows indicate the median sSFR value of each distribution.}
\label{fig_7}
\end{figure}

We thus present in Figure 7 histograms of the sSFR distribution of the samples. We notice that  spirals in elliptically dominated groups have a median sSFR that is lower than in spiral-rich groups. However, when we examine their distribution in detail, we see that they have a median of $\sim$1.29$^{+3.86}_{-3.47}$$\cdot10^{-11}$ yr$^{-1}$, while galaxies in spiral-rich groups have a median of  $\sim$2.71$^{+2.45}_{-1.29}$$\cdot10^{-11}$ yr$^{-1}$. As explained above, this minor difference could be attributed to the possibility that, in dynamically ``old" groups, spiral galaxies have already interacted several times in the past with the other members of their groups and as a result had more time to increase their stellar mass. Also their SFR is much lower than in the past, since they have consumed most of their gas. These galaxies are located in the lower left part of Lacy's plot.

It is interesting to note that the sSFR of spirals in spiral-rich groups  is similar to the sSFR of interacting pair spirals. 
This appears to contrast to the fact that galaxies in compact groups do experience more interactions from their close neighbors, either major or minor mergers, than a single galaxy by its companion. A possible explanation could be that the higher velocity dispersion of compact groups \citep[$\sim$330 km s$^{-1}$][]{Hickson92}, compared to typical values seen in near parabolic prograde encounters of galaxy pairs, reduces the gravitational impulse exerted on each galaxy. Furthermore,  the complex geometry and orbits of the group members could strip some of the gas out of these galaxies rather than funnel it to the central part, thereby increasing the SFR. This is consistent with the observed morphology of the group members, since $\sim43\%$ of them display features such as tidal tails, bridges, etc. \citep{Mendes94}. Another explanation for this similarity could be that spiral-rich compact groups are simply chance alignments of interacting pairs \citep{Mamon08}. In this scenario, two or more interacting galaxy pairs at different redshifts have a projected separation that makes them appear as a compact group. However, this is not the case for our sample, as only real group members were considered in our analysis.

\subsection{Cold dust in Hickson compact groups}

For estimating the total amount of dust in a galaxy, far-IR observations are essential, since the bulk of the dust mass in galaxies is at low temperatures ($<20$K). 
The only far-IR data available for our sample are based on IRAS \citep[see][]{Allam96}, since the Akari all-sky maps have not yet been released. 
However, the IRAS data suffer from poor spatial resolution and in nearly all 
cases one cannot resolve the emission of individual group members.  Even though there is no high 
spatial-resolution, far-IR imagery for groups, one would expect that gas and dust could be stripped from 
the group members during their tidal interactions. This material could remain outside the areas of 
the galaxies and  would have low temperature \citep[see][]{Xilouris06}. Such a case has been detected in the nearby giant 
elliptical galaxy Centaurus A, where $\sim$15K dust has been detected outside the main body of the 
galaxy \citep[see][]{Stickel04}. This was understood as a consequence of the accretion by a number of gas-rich 
dwarf companion galaxies that surround Centaurus A and have led to the formation of its dust lane, as 
well as to the filaments of atomic and molecular gas that surrounds it \citep[see][]{Schiminovich94, Charmandaris00}. Based on our Spitzer data, warm dust in the intragroup environment has been seen in some groups (HCG40, 54, 55, and 79) in the form of  diffuse mid-IR light or an extended halo around groups members. 

\begin{figure}
\resizebox{\hsize}{!}{\includegraphics[scale=1.0]{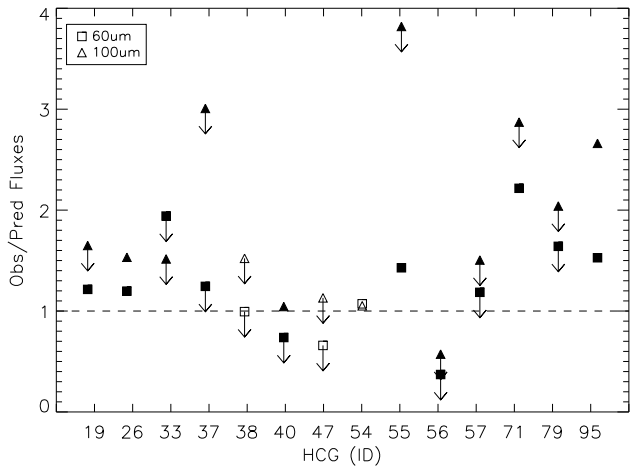}}
\caption{Plot of the ratio of the observed IRAS 60 and 100 $\mu$m fluxes of each group divided by the corresponding flux predicted by our SED fits. Since IRAS did not resolve/detect all the members of each group, several of the ratios are indicated as upper limits. Filled symbols indicate spiral-poor groups.}
\label{fig_1}
\end{figure}

Using the mid-IR imaging and extrapolating to the far-IR by the SED fits in this work, we can estimate the fraction of  the far-infrared luminosity that is contributed by each group member to the total far-IR luminosity of the whole group. We can calculate the ``synthetic'' IRAS 60$\mu$m and 100$\mu$m flux densities of each group and compare them to the ones deduced by \citet{Allam96}. The results are presented in Table 3 and Figure 8. In most cases, IRAS could not detect all the members of each group, so upper limits were used for then. Consequently, our total flux estimate is an upper limit. We observe that there are 9 groups, HCG19, HCG26, HCG33, HCG37, HCG55, HCG57, HCG71, HCG79, and HCG95, where the emission detected by IRAS is clearly higher than the total emission estimated using our galaxy SED fits. Most of these groups display evidence of strong tidal encounters and merging. This agrees with half or more of their members are early-type or peculiar galaxies and are classified as dynamically ``old'',  while only  one (HCG55) has the form of a compact chain. It is thus likely that these groups have diffuse cold dust in the intragroup region. Part of this excess could also be caused by the uncertainties introduced by our SED fits when we extrapolated to the 60 and 100$\mu$m flux of the galaxies. This uncertainty is real and can be seen if we notice that, in most groups, the $100\mu$m flux ratio is higher than the 60$\mu$m one, possibly because it is farther away than our last anchor point at $24\mu$m. A more detailed analysis of the SEDs from the UV to the mid-IR for all these groups using the models developed by \cite{daCunha08}, along with observations with PACS and SPIRE instrument onboard the Herschel Space Telescope, will address this issue conclusively.

\section{Conclusions}

In this paper we have presented our first analysis of the near- and mid-infrared SED of 69 galaxies contained in 14 Hickson compact groups, and found the following:

- Nearly half of the galaxies in the groups, 14 out of the 32, which are optically classified  as ellipticals, have mid-IR emission and colors  consistent with those expected for late-type systems. We suggest that this stems from enhanced star formation as a result of recent gas accretion from companion galaxies in the groups.

- Based on their integrated mid-IR color, 16 galaxies (23\%) are AGN candidates. Detailed nuclear photometry reveals that 6 of these have nuclear mid-IR SEDs, which are inconsistent with a mid-IR spectrum dominated by star formation.

- We find no evidence of the SFR and build up of stellar mass, which would substantially differentiate late-type galaxies in groups from galaxies in early-stage interacting pairs, or spiral galaxies in the field. This is a surprise given that the  group environment has played an important role in the evolution of the galaxies, shown by the fact that most groups contain a large fraction of early-type systems. However, late-type galaxies in dynamically ``old'' HCGs might have a lower, though not statistically significant,  sSFR than those in dynamically ``young'' groups, which could be attributed to multiple past interactions. 

- We investigated the contribution of each galaxy to the total far-IR emission of its group, and identify 9 groups where extended cold dust emission may be present. 

\begin{acknowledgements}  
We would like to thank E. da Cunha, G. Magdis and D. Elbaz for useful discussions, an anonymous referee for the detailed comments that improved this paper. TB, VC, and TDS,  would like to acknoledge partial support from the EU ToK grant 39965 and FP7-REGPOT 206469.
\end{acknowledgements} 





\begin{center}
\begin{table*}
\caption{Near-IR and mid-IR photometry of our Hickson compact group sample}
\label{table:1}
\begin{tabular}{cccccccccccc}
\hline
\hline
HCG & Type & z & J & H & Ks & $3.6\mu$m & $4.5\mu$m & $5.8\mu$m & $8.0\mu$m & $12.0\mu$m & $24.0\mu$m \\ 
(ID) &  &  & (mJy) & (mJy) & (mJy) & (mJy) & (mJy) & (mJy) & (mJy) & (mJy) & (mJy) \\ 
\hline
19a & E2 & 0.014 & 43.7 & 43.4 & 33.0 & 21.3 & 13.5 & 7.5 & 4.3 & 3.8 & 1.8 \\ 
19b & Scd & 0.014 & 10.9 & 11.1 & 8.6 & 5.9 & 4.2 & 8.6 & 20.0 & 11.7 & 24.2 \\ 
19c & Sdm & 0.014 & 4.4 & 4.1 & 2.5 & 3.2 & 1.4 & 2.3 & 5.1 & ... & 6.7 \\ 
&   &   &   &   &   &   &   &   &   &   &   \\ 
26a & Scd & 0.032 & 12.8 & 13.9 & 23.0 & 12.1 & 8.5 & 16.8 & 43.5 & 12.5 & 29.2 \\ 
26b & E0 & 0.031 & 10.1 & 11.0 & 11.0 & 5.6 & 3.6 & 2.7 & 3.3 & ... & 9.0 \\ 
26c & S0 & 0.032 & 4.8 & 5.3 & 4.6 & 2.3 & 1.5 & 0.9 & 1.0 & ... & 0.9 \\ 
26d & cI & 0.030 & 1.9 & 2.2 & 2.5 & 1.2 & 0.9 & 0.6 & 0.8 & ... & ... \\ 
26e & Im & 0.032 & 0.5 & 0.6 & 1.1 & 0.7 & 0.5 & 0.9 & 2.6 & ... & 4.0 \\ 
26f & cI & 0.032 & 0.3 & 0.2 & 0.4 & 0.2 & 0.1 & ... & ... & ... & 0.3 \\ 
26g & S0 & 0.031 & 0.9 & 1.1 & 1.3 & 0.7 & 0.6 & 0.3 & 0.2 & ... & ... \\ 
 &   &   &   &   &   &   &   &   &   &   &   \\ 
33a & E1 & 0.025 & 41.7 & 45.5 & 36.5 & 27.6 & 17.9 & 10.3 & 6.6 & ... & 2.6 \\ 
33b & E4 & 0.026 & 40.9 & 44.0 & 34.9 & 24.8 & 15.5 & 9.4 & 7.1 & ... & 2.3 \\ 
33c & Sdm & 0.026 & 29.6 & 25.0 & 16.8 & 21.2 & 15.2 & 21.4 & 58.6 & 16.3 & 41.2 \\ 
33d & E0 & 0.026 & 12.1 & 13.0 & 9.7 & 5.5 & 4.0 & 1.7 & 0.9 & ... & 0.1 \\ 
 &   &   &   &   &   &   &   &   &   &   &   \\ 
37a & E7 & 0.022 & 64.7 & 79.6 & 72.1 & 61.7 & 37.2 & 19.3 & 13.1 & ... & 4.8 \\ 
37b & Sbc & 0.022 & 33.2 & 43.3 & 43.6 & 22.9 & 14.4 & 14.5 & 29.2 & 14.1 & 36.7 \\ 
37c & S0a & 0.024 & 11.1 & 11.9 & 10.4 & 3.5 & 2.3 & 1.8 & 2.1 & ... & 3.5 \\ 
37d & SBdm & 0.020 & 2.7 & 3.5 & 3.2 & 1.8 & 1.1 & 2.0 & 5.2 & 3.3 & 4.9 \\ 
37e & E0 & 0.021 & 4.4 & 5.6 & 5.2 & 2.4 & 1.5 & 1.0 & 1.1 & .. & 1.8 \\ 
 &   &   &   &   &   &   &   &   &   &   &   \\ 
38a & Sbc & 0.029 & 17.1 & 21.5 & 18.1 & 11.1 & 8.0 & 16.4 & 46.9 & ... & 44.5 \\ 
38b & SBd & 0.029 & 15.4 & 18.8 & 14.7 & 12.3 & 8.5 & 16.3 & 27.6 & 8.3 & 45.2 \\ 
38c & Im & 0.029 & 9.2 & 11.1 & 8.4 & 6.5 & 4.4 & 8.4 & 24.2 & 5.6 & 67.8 \\ 
38d & SBa & 0.029 & 6.7 & 7.8 & 6.6 & 3.4 & 2.5 & 1.2 & 1.4 & ... & 0.4 \\ 
 &   &   &   &   &   &   &   &   &   &   &   \\ 
40a & E3 & 0.022 & 79.4 & 78.7 & 74.2 & 45.4 & 28.0 & 20.2 & 15.3 & ... & 4.0 \\ 
40b & S0 & 0.023 & 35.1 & 33.3 & 31.1 & 18.7 & 12.2 & 8.9 & 8.8 & ... & 3.0 \\ 
40c & Sbc & 0.021 & 37.3 & 43.1 & 45.7 & 29.4 & 19.7 & 34.7 & 83.5 & ... & 73.4 \\ 
40d & SBa & 0.022 & 31.7 & 32.0 & 30.7 & 21.3 & 14.3 & 22.7 & 66.9 & ... & 88.1 \\ 
40e & Sc & 0.022 & 8.8 & 8.7 & 8.5 & 6.4 & 4.1 & 4.0 & 5.9 & ... & 5.6 \\ 
40f & E1 & 0.021 & 5.9 & 4.4 & 3.0 & 1.4 & 1.0 & 0.9 & 0.8 & ... & ... \\ 
 &   &   &   &   &   &   &   &   &   &   &   \\ 
47a & Sbc & 0.032 & 26.8 & 35.0 & 28.0 & 20.7 & 13.3 & 20.3 & 52.3 & 22.1 & 87.0 \\ 
47b & E3 & 0.032 & 16.0 & 21.4 & 16.9 & 11.8 & 7.2 & 4.1 & 2.4 & ... & 0.6 \\ 
47c & Sc & 0.032 & 4.5 & 5.9 & 4.5 & 3.0 & 1.9 & 2.3 & 2.4 & ... & 9.9 \\ 
47d & Sd & 0.032 & 5.7 & 7.1 & 5.9 & 3.6 & 2.4 & 2.3 & 6.4 & ... & 4.5 \\ 
 &   &   &   &   &   &   &   &   &   &   &   \\ 
54a & Sdm & 0.005 & 10.2 & 8.6 & 6.0 & 6.7 & 5.3 & 6.8 & 10.3 & ... & 4.4 \\ 
54b & Im & 0.005 & 5.6 & 5.6 & 4.4 & 3.7 & 2.9 & 5.5 & 11.1 & ... & 22.7 \\ 
54c & Im & 0.005 & 1.3 & 1.1 & 1.4 & 2.3 & 1.5 & 2.1 & 4.1 & ... & 2.8 \\ 
54d & Im & 0.006 & 0.5 & 0.7 & 0.6 & 0.5 & 0.4 & 0.3 & 0.7 & ... & 0.7 \\ 
 &   &   &   &   &   &   &   &   &   &   &   \\ 
55a & E0 & 0.054 & 14.4 & 17.2 & 16.6 & 6.5 & 3.9 & 2.7 & 1.2 & ... & 0.6 \\ 
55b & S0 & 0.052 & 7.3 & 7.4 & 7.3 & 3.4 & 2.2 & 1.3 & 0.7 & ... & 0.3 \\ 
55c & E3 & 0.052 & 7.7 & 10.8 & 11.6 & 5.5 & 3.8 & 3.9 & 10.9 & ... & 7.7 \\ 
55d & E2 & 0.054 & 4.3 & 4.9 & 4.8 & 2.1 & 1.4 & 0.9 & 0.7 & ... & 0.6 \\ 
55e & Sc & 0.054 & 1.2 & 2.2 & 2.2 & 1.8 & 1.3 & 0.9 & 4.0 & ... & 6.3 \\ 
 &   &   &   &   &   &   &   &   &   &   &   \\ 
56a & Sc & 0.027 & 15.7 & 25.0 & 13.5 & 6.8 & 4.4 & 6.1 & 13.0 & 3.9 & 13.4 \\ 
56b & SB0 & 0.026 & 35.0 & 52.9 & 29.5 & 25.9 & 30.4 & 39.0 & 53.7 & 49.8 & 185 \\ 
56c & S0 & 0.027 & 18.8 & 24.8 & 15.0 & 7.7 & 3.3 & 3.8 & 3.2 & ... & 0.9 \\ 
56d & S0 & 0.028 & 9.3 & 13.8 & 7.8 & 4.7 & 3.1 & 5.9 & 14.8 & 5.4 & 20.2 \\ 
56e & S0 & 0.027 & 5.7 & 7.2 & 4.4 & 2.0 & 1.4 & 2.1 & 4.2 & 1.8 & 6.6 \\ 
 &   &   &   &   &   &   &   &   &   &   &   \\ 
57a & Sbc & 0.029 & 56.7 & 62.2 & 67.9 & 33.8 & 18.5 & 15.0 & 18.1 & 12.5 & 13.8 \\ 
57b & SBb & 0.030 &26.7 & 26.8 & 27.2 & 14.1 & 8.9 & 6.8 & 10.0 & ... & 8.1 \\ 
57c & E3 & 0.030 & 22.1 & 23.0 & 23.7 & 12.8 & 8.7 & 4.2 & 2.7 & 1.4 & 0.4 \\ 
57d & SBc & 0.030 & 13.1 & 13.4 & 14.1 & 7.6 & 4.9 & 7.4 & 18.1 & 6.6 & 24.0 \\ 
57e & S0a & 0.030 & 20.1 & 20.8 & 21.4 & 9.5 & 6.1 & 3.8 & 4.1 & ... & 2.1 \\ 
57f & E4 & 0.031 & 13.7 & 13.6 & 14.3 & 6.7 & 4.3 & 1.5 & 1.0 & ... & 0.9 \\ 
57g & SB0 & 0.032 & 11.7 & 11.8 & 12.0 & 5.4 & 3.3 & 1.5 & 0.7 & ... & ... \\ 
57h & SBb & 0.031 & 3.8 & 3.9 & 3.9 & 1.8 & 1.2 & 0.8 & 1.9 & ... & 2.5 \\ \hline
\end{tabular}
\end{table*}
\end{center}

\begin{center}
\begin{table*}
\label{table:1}
\begin{tabular}{cccccccccccc}
\hline
\hline
HCG & Type & z & J & H & Ks & $3.6\mu$m & $4.5\mu$m & $5.8\mu$m & $8.0\mu$m & $12.0\mu$m & $24.0\mu$m \\
(ID) & & & (mJy) & (mJy) & (mJy) & (mJy) & (mJy) & (mJy) & (mJy) & (mJy) & (mJy) \\ 
\hline
71a & Sbc & 0.031 & 29.9 & 36.5 & 22.6 & 17.4 & 11.3 & 16.5 & 32.6 & 24.0 & 29.2 \\ 
71b & SB0 & 0.032 & 17.3 & 21.2 & 14.1 & 8.5 & 5.8 & 10.6 & 9.1 & 9.3 & 52.7 \\ 
71c & Sbc & 0.029 & 3.5 & 4.4 & 2.8 & 2.0 & 1.2 & 2.1 & 5.2 & ... & 4.3 \\ 
71d & SB0 & 0.031 & 6.3 & 7.7 & 5.1 & 3.2 & 2.1 & 1.2 & 1.1 & ... & 2.5 \\ 
 &   &   &   &   &   &   &   &   &   &   &   \\ 
79a & Sa & 0.015 & 33.5 & 36.0 & 27.4 & 23.7 & 15.2 & 18.8 & 30.9 & 11.9 & 23.3 \\ 
79b & S0 & 0.015 & 37.3 & 40.3 & 30.9 & 17.6 & 14.6 & 15.8 & 12.2 & 6.6 & 33.7 \\ 
79c & S0 & 0.014 & 12.6 & 12.7 & 9.1 & 10.2 & 5.7 & 4.0 & 2.0 & ... & 1.2 \\ 
79d & Sdm & 0.015 & 4.3 & 3.2 & 3.2 & 2.2 & 1.2 & 1.5 & 3.5 & ... & 5.2 \\ 
79e & Scd & 0.015 & 5.7 & 6.1 & 5.3 & 4.3 & 3.3 & 4.0 & 17.5 & 5.8 & 0.5 \\ 
 &   &   &   &   &   &   &   &   &   &   &   \\ 
95a & Sbc & 0.040 & 30.5 & 33.5 & 28.6 & 18.0 & 11.0 & 6.7 & 7.8 & ... & 1.3 \\ 
95b & SB0 & 0.039 & 11.9 & 13.0 & 13.3 & 9.2 & 6.4 & 14.1 & 30.4 & 23.1 & 41.2 \\ 
95c & Sbc & 0.039 & 10.4 & 11.1 & 9.7 & 7.1 & 4.3 & 4.9 & 10.6 & 13.6 & 19.8 \\ 
95d & SB0 & 0.041 & 7.7 & 8.5 & 9.4 & 4.8 & 3.1 & 2.6 & 6.7 & 4.7 & 5.6\\
\hline
\end{tabular}
\end{table*}
\end{center}

\pagebreak

\begin{center}
\begin{table*}
\caption{Stellar mass, IR luminosity, Star Formation Rate, and specific SFRs for the sample}
\label{table:2}
\begin{tabular}{cccccc}
\hline \hline
HCG & log(M) & log(L$_{\rm IR}$) & SFR & sSFR  \\ 
(ID) & M$_{\odot}$ & L$_{\odot}$ & M$_{\odot}$ yr$^{-1}$ & $10^{-11}$ yr$^{-1}$ \\ 
\hline
19a & 10.57 & ... &... & ... \\ 
19b & 9.99 & 9.38 & 0.24 & 2.54 \\ 
19c & 9.46 & 8.84 & 0.08 & 2.75 \\ 
&   &   &   &   \\ 
26a & 10.91& 10.34 & 1.28 & 0.92 \\ 
26b & 10.81 & 9.65 & 0.44 & 0.68 \\ 
26c & 10.45 & ... & ... & ... \\ 
26d & 10.12 & 8.73 & 0.08 & 0.40 \\ 
26e & 9.83 & 9.46 & 0.23 & 3.34 \\ 
26f & 9.12 & 8.49 & 0.02 & 0.85 \\ 
26g & 9.87 & ... & ... & ... \\ 
 &   &   &   &   \\ 
33a & 11.13& ... & ... & ... \\ 
33b & 11.15& ... & ... & ... \\ 
33c & 10.93 & 10.18 & 1.21 & 1.79 \\ 
33d & 10.59 & ... & ... & ... \\ 
 &   &   &   &   \\ 
37a & 11.32& ... & ... & ... \\ 
37b & 11.10& 9.91 & 0.81 & 0.65 \\ 
37c & 10.55 & 9.23 & 0.12 & 0.33 \\ 
37d & 9.88 & 9.09 & 0.12 & 1.52 \\ 
37e & 10.13 & 8.60 & 0.05 & 0.38 \\ 
 &   &   &   &   \\ 
38a & 10.96& 10.42 & 1.55 & 1.73 \\ 
38b & 10.87 & 10.21 & 1.57 & 2.17 \\ 
38c & 10.63 & 10.37 & 2.25 & 5.40 \\ 
38d & 10.55 & ... & ... & ... \\ 
&   &   &   &   \\ 
40a & 11.33& ... & ... & ... \\ 
40b & 10.99& ... & ... & ... \\ 
40c & 11.08& 10.34 & 1.35 & 1.15 \\ 
40d & 10.95& 10.41 & 1.78 & 2.00 \\ 
40e & 10.39 & 9.25 & 0.16 & 0.63 \\ 
40f & 9.89 & ... & ... & ... \\ 
 &   &   &   &   \\ 
47a & 11.21& 10.59 & 3.19 & 2.00 \\ 
47b & 11.02& ... & ... & ... \\ 
47c & 10.45 & 9.72 & 0.51 & 1.81 \\ 
47d & 10.57 & 9.49 & 0.26 & 0.68 \\ 
 &   &   &   &   \\ 
54a & 9.09 & 7.97 & 0.01 & 1.02 \\ 
54b & 8.79 & 8.46 & 0.04 & 5.99 \\ 
54c & 8.33 & 7.81 & 0.01 & 3.49 \\ 
54d & 8.09 & 7.37 & 0.002 & 1.91 \\ 
 &   &   &   &   \\ 
55a & 11.37& 8.92 & 0.12 & 0.04 \\ 
55b & 11.10& 8.59 & 0.06 & 0.04 \\ 
55c & 11.30& 10.18 & 1.05 & 0.50 \\ 
55d & 10.95& 8.99 & 0.12 & 0.12 \\ 
55e & 10.64 & 10.02 & 0.91 & 2.18 \\ 
 &   &   &   &   \\ 
56a & 10.77& 9.68 & 0.48 & 0.82 \\ 
56b & 11.08& 10.71 & 4.44 & 3.85 \\ 
56c & 10.82 & ... & ... & ... \\ 
56d & 10.56 & 9.86 & 0.73 & 2.02 \\ 
56e & 10.28 & 9.18 & 0.26 & 1.35 \\ 
 &   &   &   &   \\ 
57a & 11.54& 9.99 & 0.58 & 0.16 \\ 
57b & 11.14& 9.69 & 0.39 & 0.26 \\ 
57c & 11.11& ... & ... & ... \\ 
57d & 10.88 & 9.96 & 0.97 & 1.28 \\ 
57e & 11.07& 9.17 & 0.11 & 0.10 \\ 
57f & 10.92& ... & ... & ... \\ 
57g & 10.87 & ... & ... & ... \\ 
57h & 10.36 & 9.22 & 0.15 & 0.62 \\ \hline
\end{tabular}
\end{table*}
\end{center}

\begin{center}
\begin{table*}
\begin{tabular}{ccccc}
\hline \hline
HCG & log(Mass) & log(L$_{\rm IR}$) & SFR & sSFR  \\ 
(ID) & M$_{\odot}$ & L$_{\odot}$ & M$_{\odot}$ yr$^{-1}$ & $10^{-11}$ yr$^{-1}$ \\ 
\hline
71a & 11.11& 10.10 & 1.25 & 0.94 \\ 
71b & 10.93& 10.34 & 2.11 & 2.53 \\ 
71c & 10.16 & 9.39 & 0.21 & 1.42 \\ 
71d & 10.49 & ... & ... & ... \\ 
 &   &   &   &   \\ 
79a & 10.56 & 9.44 & 0.27 & 0.75 \\ 
79b & 10.61 & 9.36 & 0.38 & 0.93 \\ 
79c & 10.07 & ... & ... & ... \\ 
79d & 9.62 & 8.79 & 0.07 & 1.71 \\ 
79e & 9.87 & 9.37 & 0.21 & 2.65 \\ 
 &   &   &   &   \\ 
95a & 11.44& ... & ... & ... \\ 
95b & 11.09& 10.49 & 2.60 & 2.06 \\ 
95c & 10.95& 10.12 & 1.32 & 1.48 \\ 
95d & 10.99& 9.70 & 0.48 & 0.50 \\
\hline
\end{tabular}
\end{table*}
\end{center}

\pagebreak
\clearpage

\begin{center}
\begin{table}
\caption{Observed and predicted IRAS $60\mu$m and 100$\mu$m fluxes}
\begin{tabular}{ccccc}
\hline \hline
HCG & \multicolumn{2}{c}{IRAS $60\mu$m} & \multicolumn{2}{c}{IRAS $100\mu$m} \\ 
group & Observed & Predicted & Observed & Predicted \\ 
& mJy & mJy & mJy & mJy \\
\hline 
19 & 300 & 246.8 & 1110 & 673 \\ 
26 & 540 & 450.6 & 1496 & 976 \\ 
33 & $<660$ & 339.6 & $<1140$ & 752 \\ 
37 & $\sim560$ & 449.5 & $<2000$ & 665 \\ 
38 & $\sim1430$ & 1439 & $<3160$ & 2076 \\ 
40 & $<1090$ & 1478 & 3850 & 3686 \\ 
47 & $<580$ & 880.9 & $<1510$ & 1336 \\ 
54 &  300 & 278.2 & 710 & 675 \\ 
55 &  200 & 139.4 & 1630 & 322 \\ 
56 & $<690$ & 1860 & $<1510$ & 2637 \\ 
57 & $<460$ & 437.5 & $<1380$ & 917 \\ 
71 & 1640 & 736.9 & $<3070$ & 1069 \\ 
79 & $\sim960$ & 584.4 & $<2320$ & 1137 \\ 
95 &  941 & 616.2 & 2390 & 898 \\
\hline
\end{tabular}
\end{table}
\end{center}

\clearpage
\newpage

\appendix
\section{Infrared morphology}

In this section we briefly describe the  overall appearance of groups in the Spitzer mid-IR images. A detailed analysis of the 
mid-IR and near-IR color profiles of the galaxies will be the subject of a future paper.

\begin{itemize}
\item HCG19: Galaxy a is faint in the mid-IR displaying only a luminous core at $8\mu$m. Member b is very 
bright in all mid-IR bands, and its spiral arms  are clearly visible at 5.8, and 8$\mu$m, likely due to 
associated PAH emission. Finally, galaxy c has faint extended emission in all bands.

\item HCG26: This is  a very dense group. Members a, b, d, and g are merging with a and e the brightest. 
There is also a feature similar to a tidal tail over galaxy g which is visible in the optical images but not in the 
mid-IR.

\item HCG33: Galaxies a, b, and d are early-type and so are very faint in the mid-IR; however, galaxy c is classified as type Sdm 
and is very bright in all Spitzer bands.

\item HCG37: Member a displays a very bright extended halo in the first IRAC band, but it is not so bright in the other bands. Member b presents strong nuclear emission at 8 and 24 $\mu$m. Member d is also bright in these wavelengths. Finally, member c has an IR-bright nucleus, and it is spectroscopically classified as a low-luminosity AGN.

\item HCG38: Galaxies b and c form an interacting pair both displaying very luminous extended tails with bright spots, possibly because of giant HII regions. Galaxy a is also luminous in the mid-IR.

\item HCG40: There is a $3.6\mu$m extended halo around the group. Galaxies c and d are very bright, especially at 24$\mu$m. We can also clearly see the spiral arms of member e, as well as nuclear emission in a and b.

\item HCG47: Galaxy a has a very bright nucleus and spiral arms while a mid-IR luminous bridge connects it to member b, which only has faint emission away from its nucleus. Galaxy c is also bright. Finally, member d is also rather faint in the mid-IR emission.

\item HCG54: Member b is a very bright galaxy, in particular at $24\mu$m. Member a appears to have two cores. One is bright at $24\mu$m and the other is so the 5.6 and 8$\mu$m bands. 

\item HCG55: An extended halo  at $3.6\mu$m, is seen around the group. Galaxies c and e are most luminous especially at 24$\mu$m. 

\item HCG56: Member b is the brightest galaxy of this group and one of the brightest of our sample.  Galaxies a, d, and e are mid-IR luminous, while c has a bright nucleus visible only in the four IRAC bands. Finally, there is a bridge between members b and c seen only in $3.6\mu$m, suggesting that it contains mostly stars.

\item HCG57: Galaxies a and d have very luminous spirals, also presenting some mid-IR bright spots.  They also seem to have a common halo at $24\mu$m . Member b also has bright spiral arms with two bright spots on them seen at  $24\mu$m. The latter coincide with the location of two identified supernovae (SN-2002AR and SN-2005BA).

\item HCG71: Member a has a bright nucleus and very extended luminous spiral arms. The nucleus of galaxy b is bright, and c is also mid-IR bright.

\item HCG79: There is an extended halo around the group seen in the $3.6\mu$m band. Galaxies a, b, d, and e are very luminous in mid-IR. There are also a tidal tail in member b and an extended feature over galaxy c. Finally, it seems that there is a bridge between a and d.

\item HCG95: There is a bridge connecting galaxies a and c, observed in all wavelengths with bright spots on it. Members c and d are mid-IR bright. Galaxy a only has a luminous nucleus, while the spiral arms of galaxy d are seen in the IRAC images .
\end{itemize}
As seen from the above descriptions, the mid-IR morphology of the galaxies in the compact groups displays clear evidence of tidal interactions and shows the effects of the dynamically induced star formation activity in them.

\begin{figure*}
\includegraphics[scale=0.9]{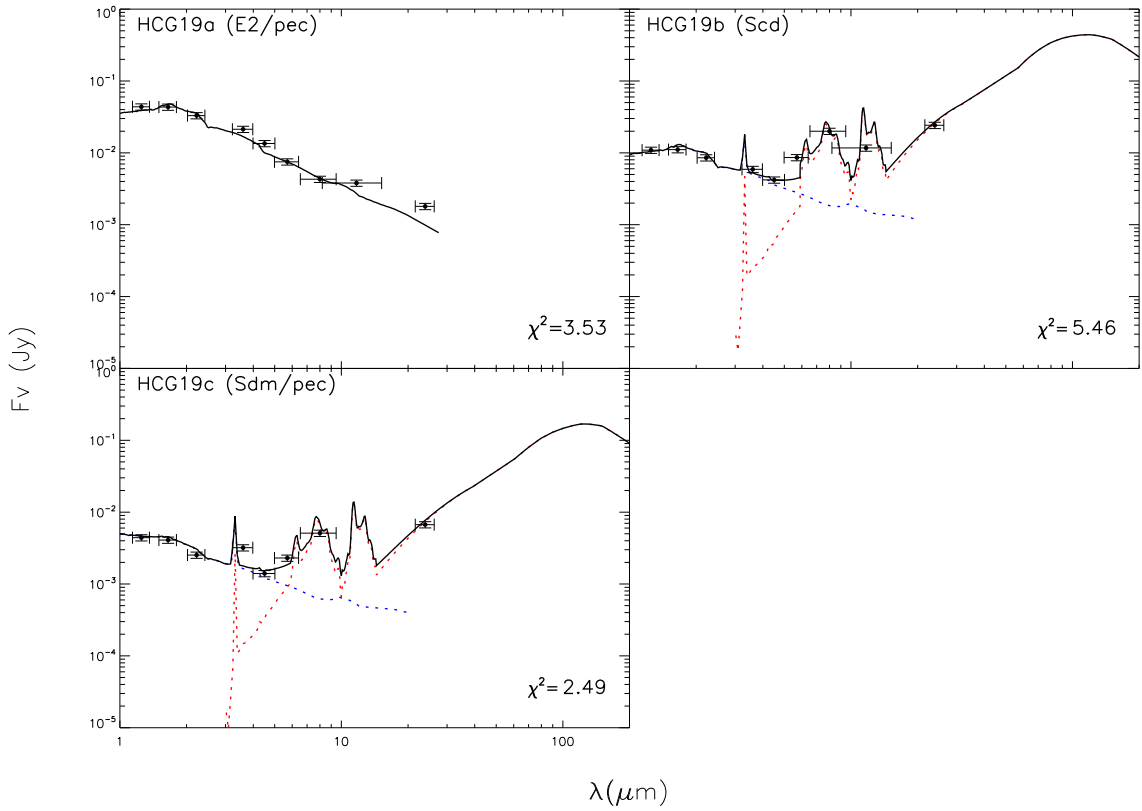}
\caption{Infrared SEDs of the HCG galaxies. The name of each galaxy and its optical classification by \citet{Hickson82} is marked at the top left of each plot. The vertical error bars show the uncertainty of the flux, while the horizontal ones indicate the bandwidth of the corresponding filter. The reduced $\chi^2$ value of each fit is marked in the bottom right of each plot.}
\end{figure*}

\begin{figure*}
\setcounter{figure}{0} 
\includegraphics[scale=0.9]{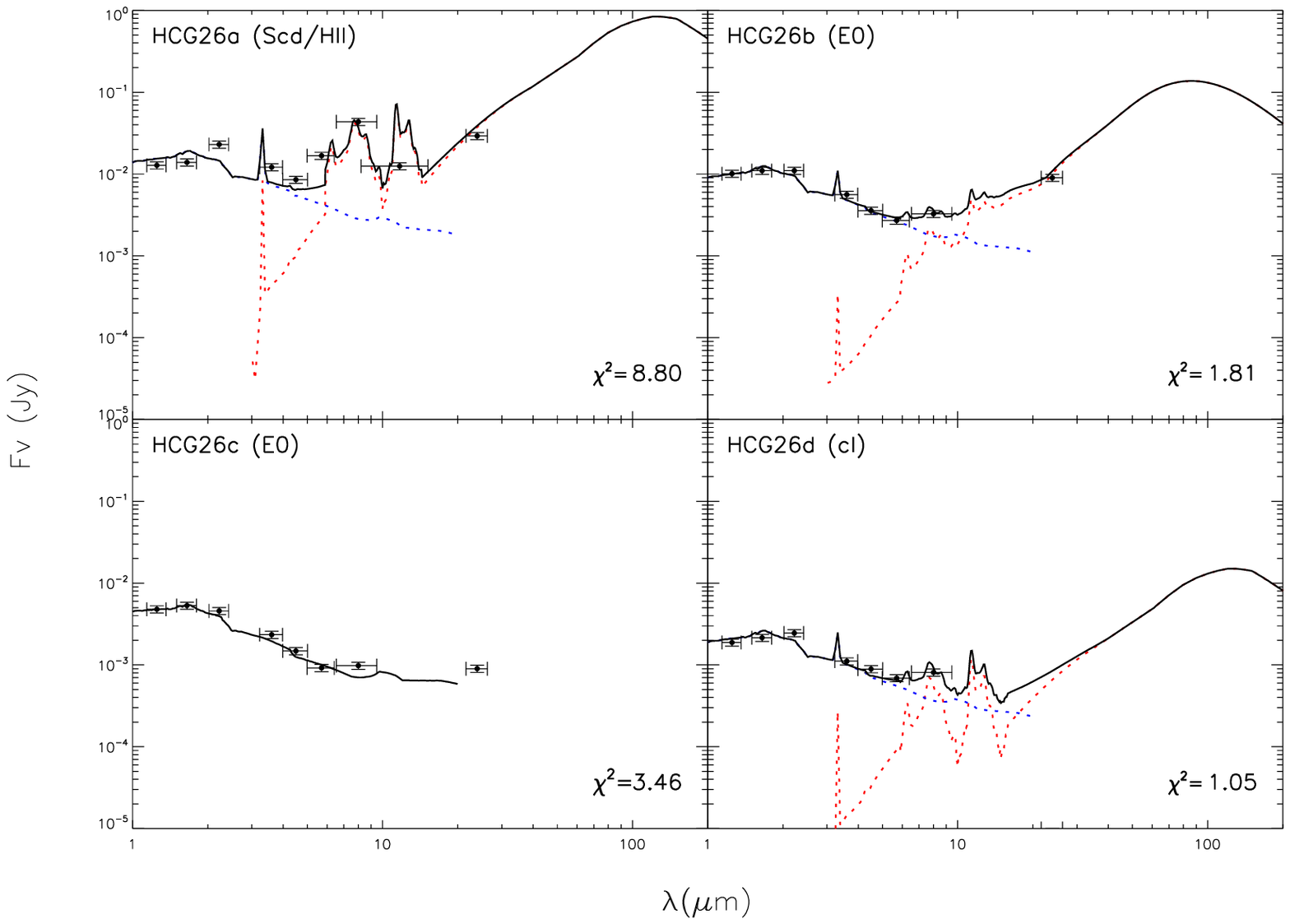}
\caption{Continued.}
\end{figure*}

\begin{figure*}
\setcounter{figure}{0} 
\includegraphics[scale=0.9]{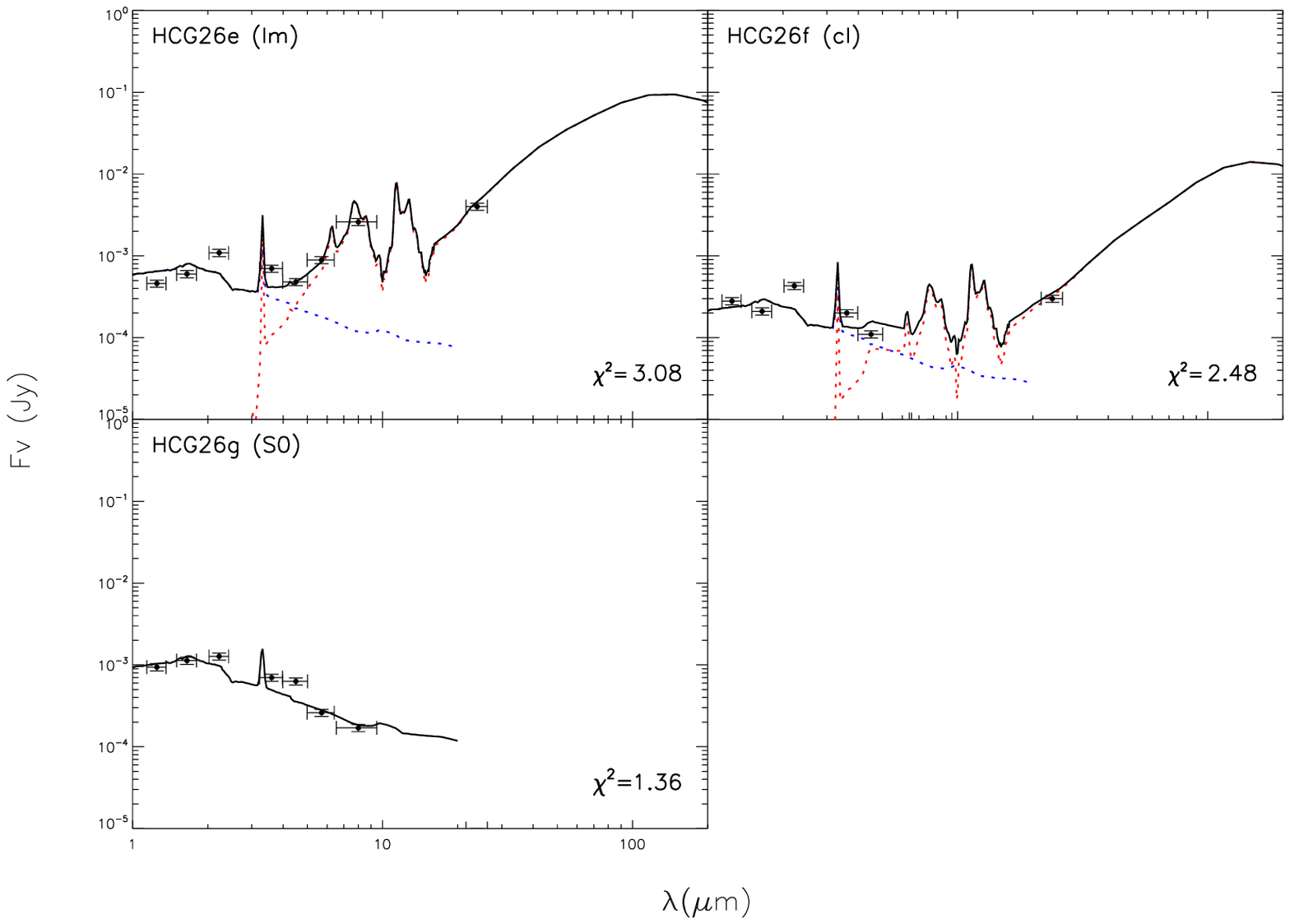}
\caption{Continued.}
\end{figure*}

\begin{figure*}
\setcounter{figure}{0} 
\includegraphics[scale=0.9]{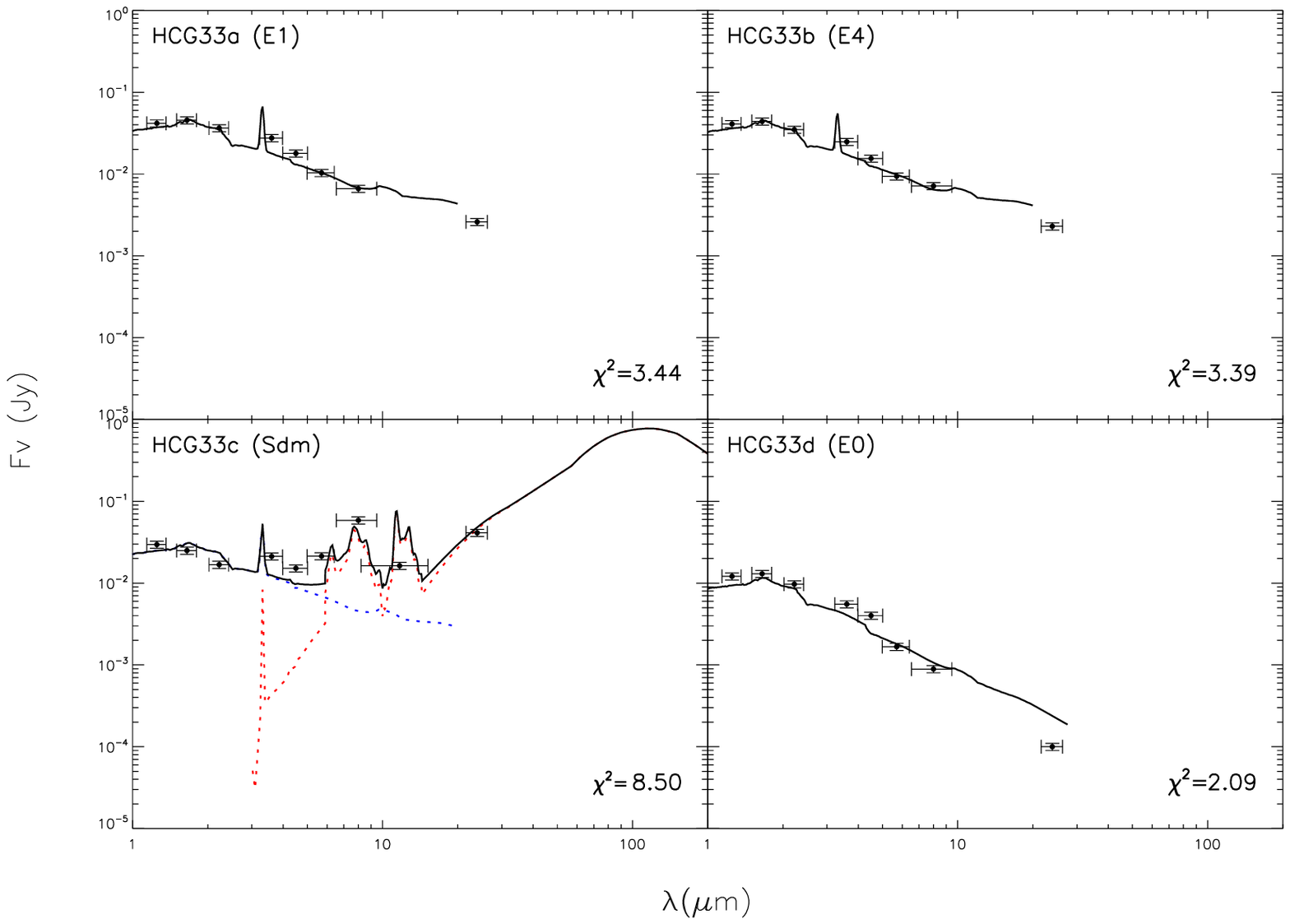}
\caption{Continued.}
\end{figure*}

\begin{figure*}
\setcounter{figure}{0} 
\includegraphics[scale=0.9]{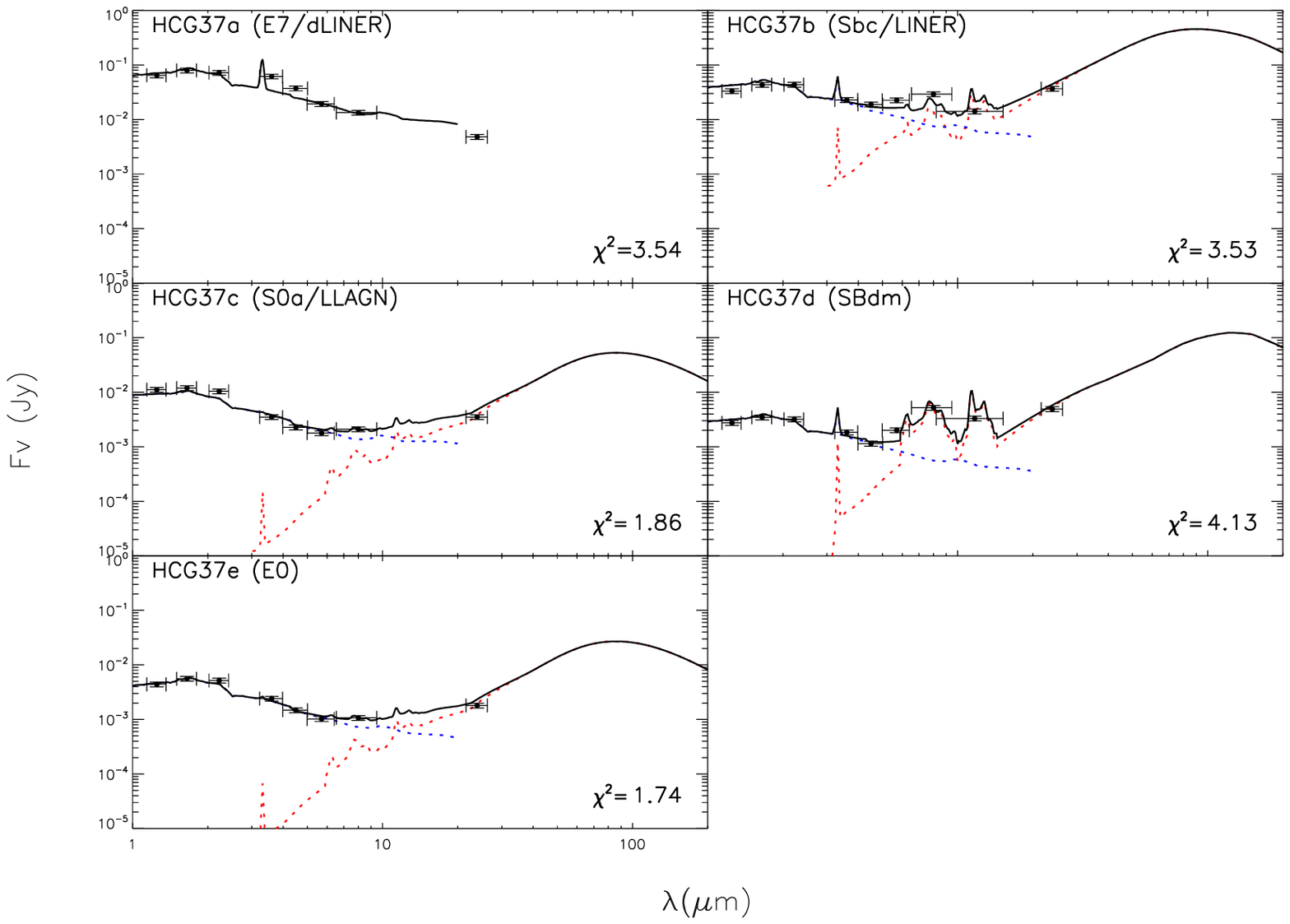}
\caption{Continued.}
\end{figure*}

\begin{figure*}
\setcounter{figure}{0} 
\includegraphics[scale=0.9]{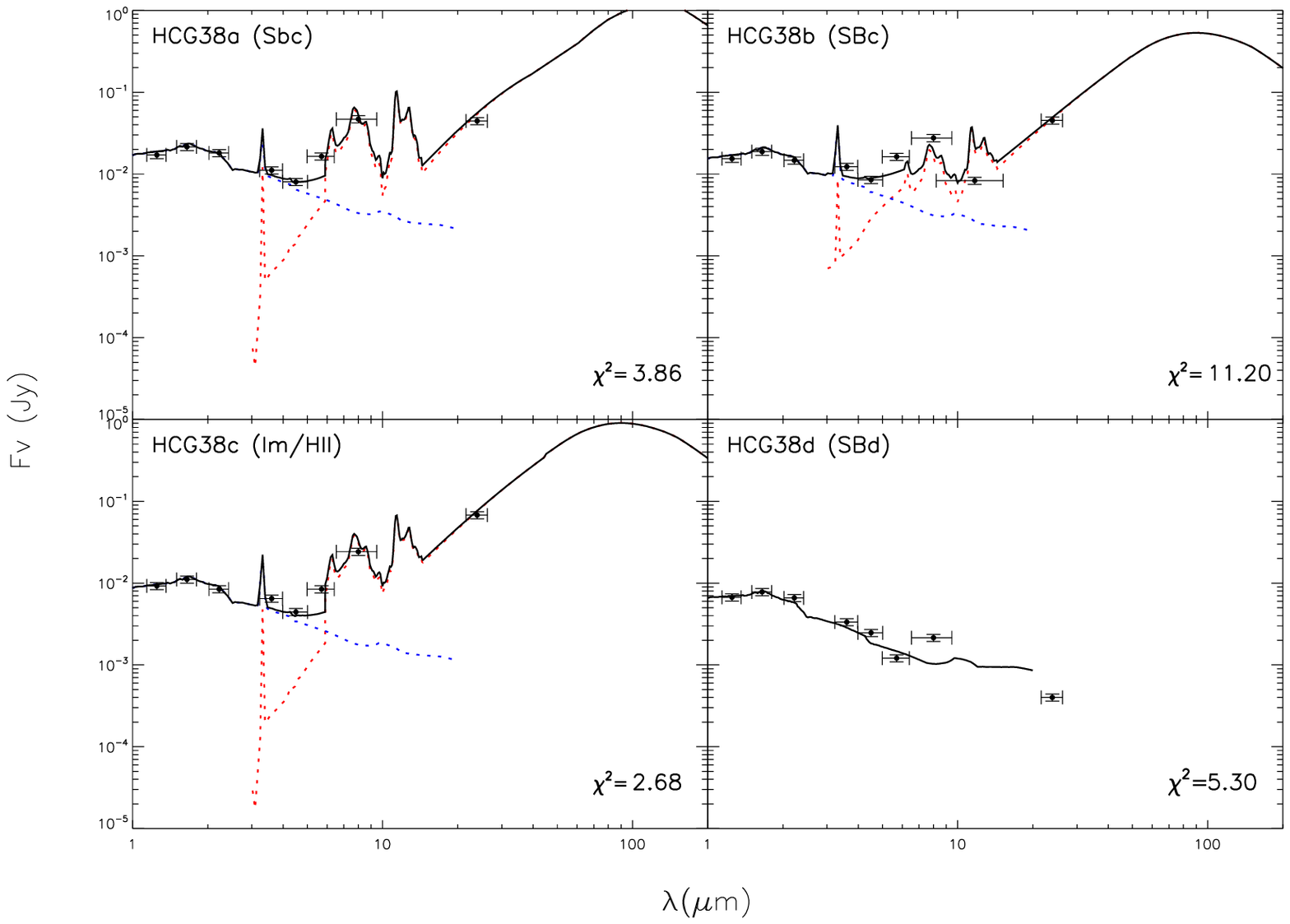}
\caption{Continued.}
\end{figure*}

\begin{figure*}
\setcounter{figure}{0} 
\includegraphics[scale=0.9]{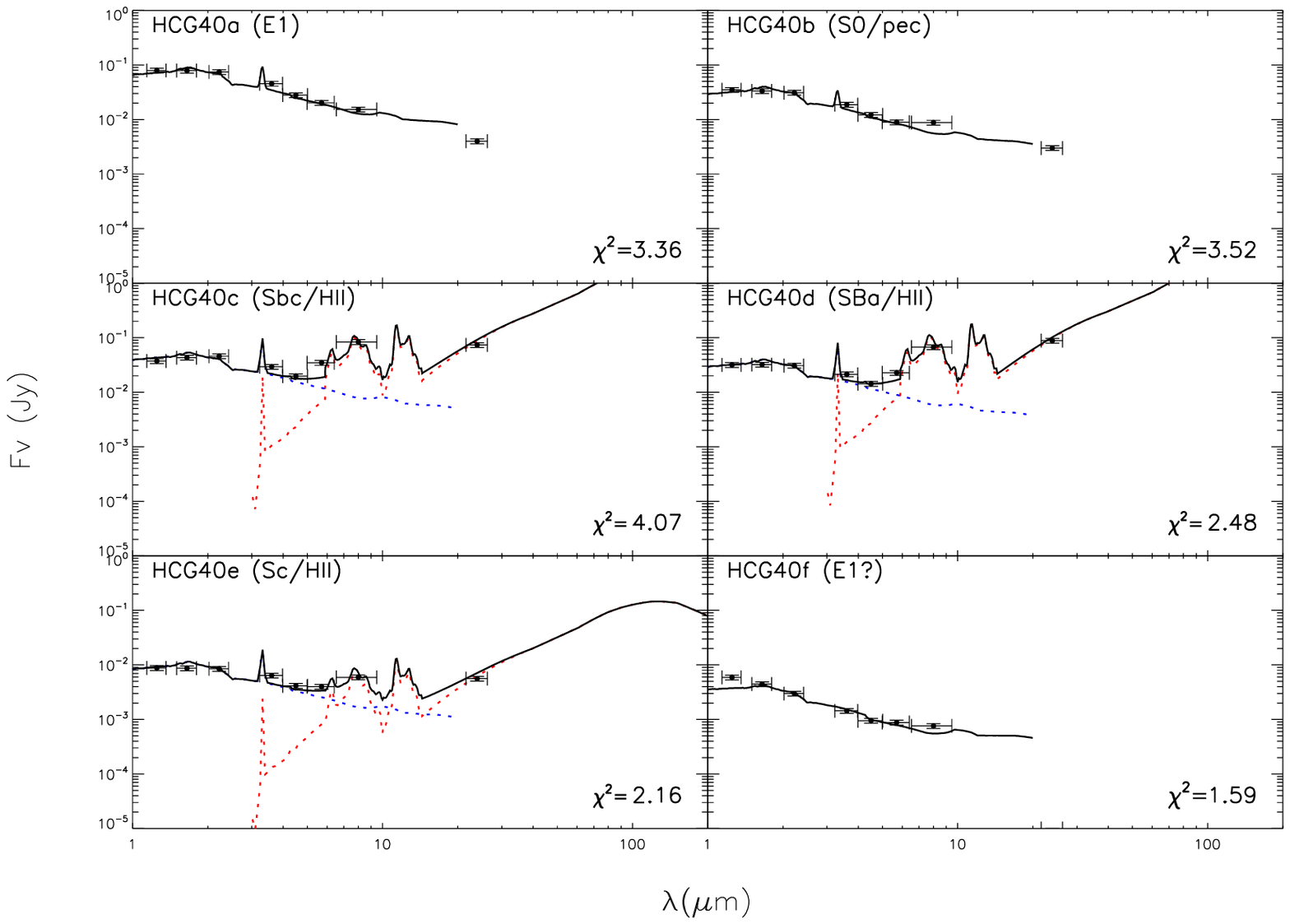}
\caption{Continued.}
\end{figure*}

\begin{figure*}
\setcounter{figure}{0} 
\includegraphics[scale=0.9]{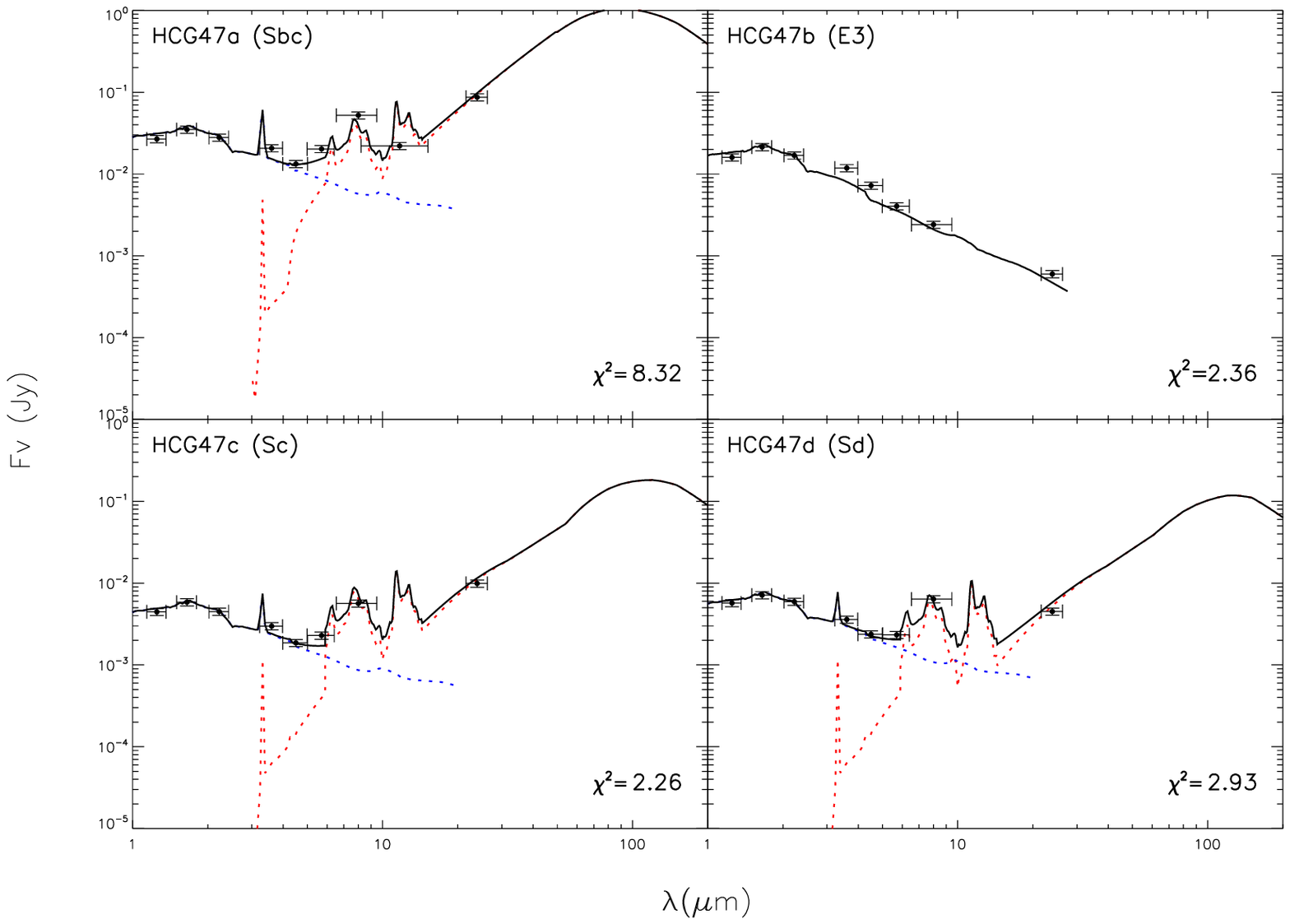}
\caption{Continued.}
\end{figure*}

\begin{figure*}
\setcounter{figure}{0} 
\includegraphics[scale=0.9]{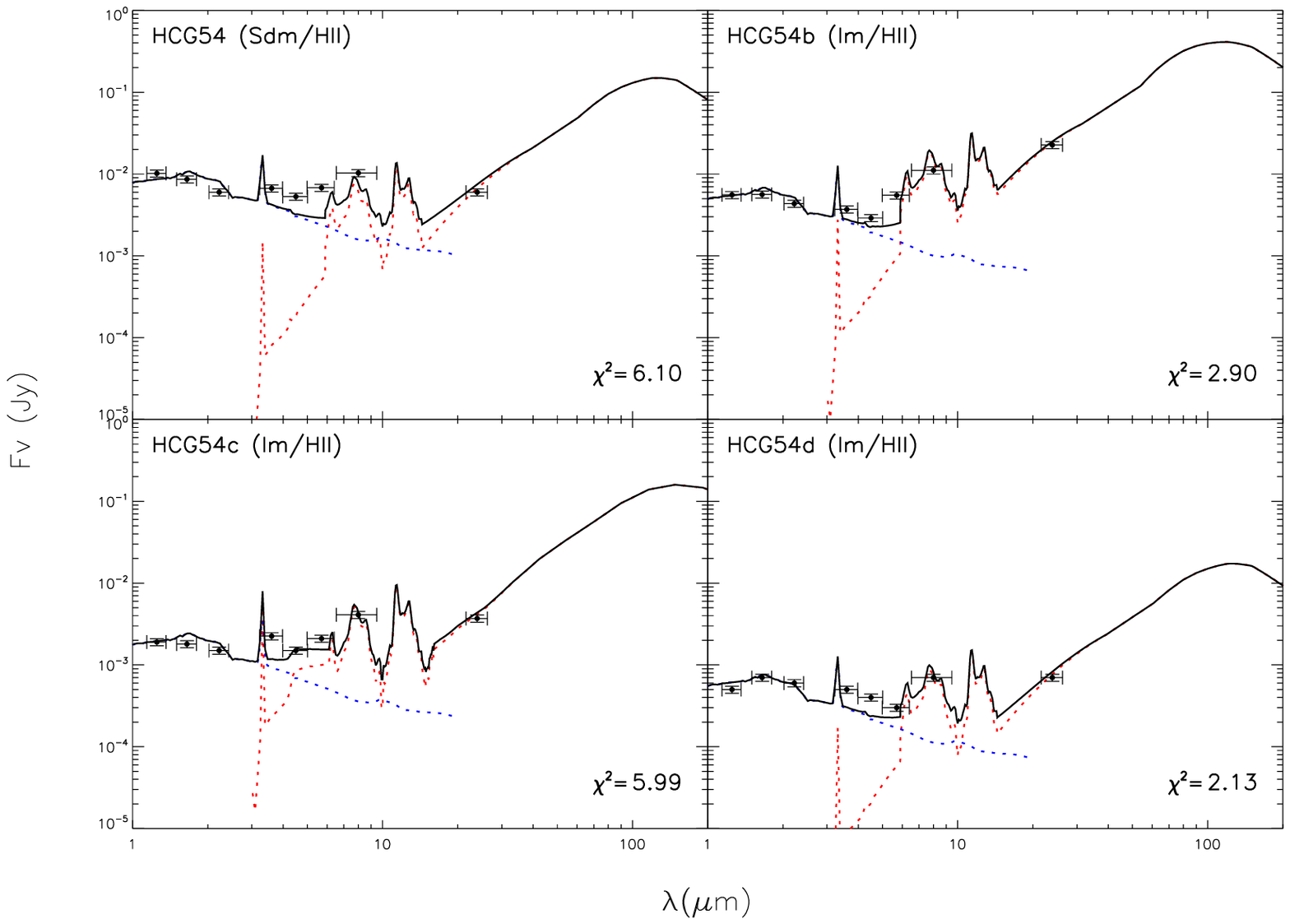}
\caption{Continued.}
\end{figure*}

\begin{figure*}
\setcounter{figure}{0} 
\includegraphics[scale=0.9]{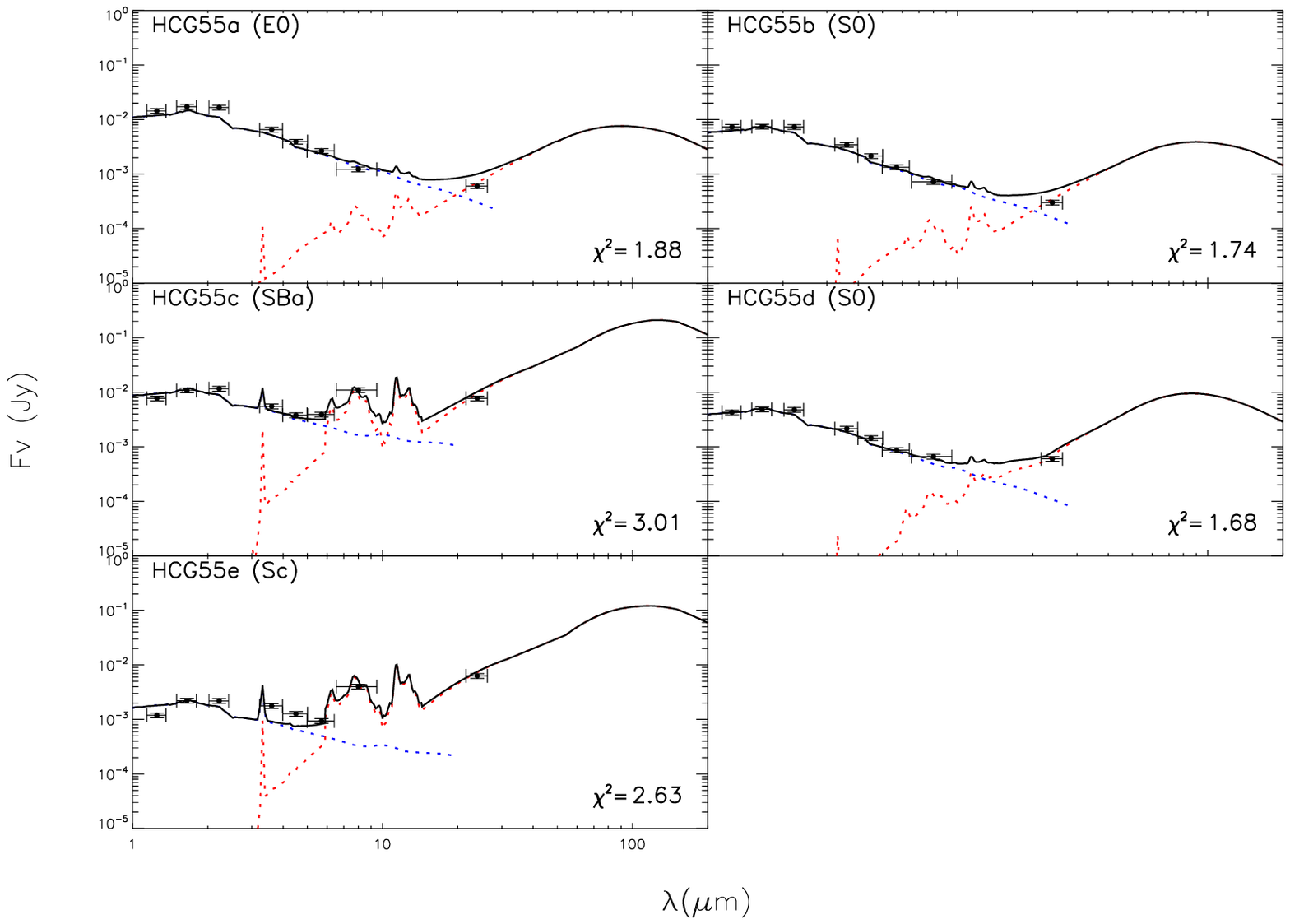}
\caption{Continued.}
\end{figure*}

\begin{figure*}
\setcounter{figure}{0} 
\includegraphics[scale=0.9]{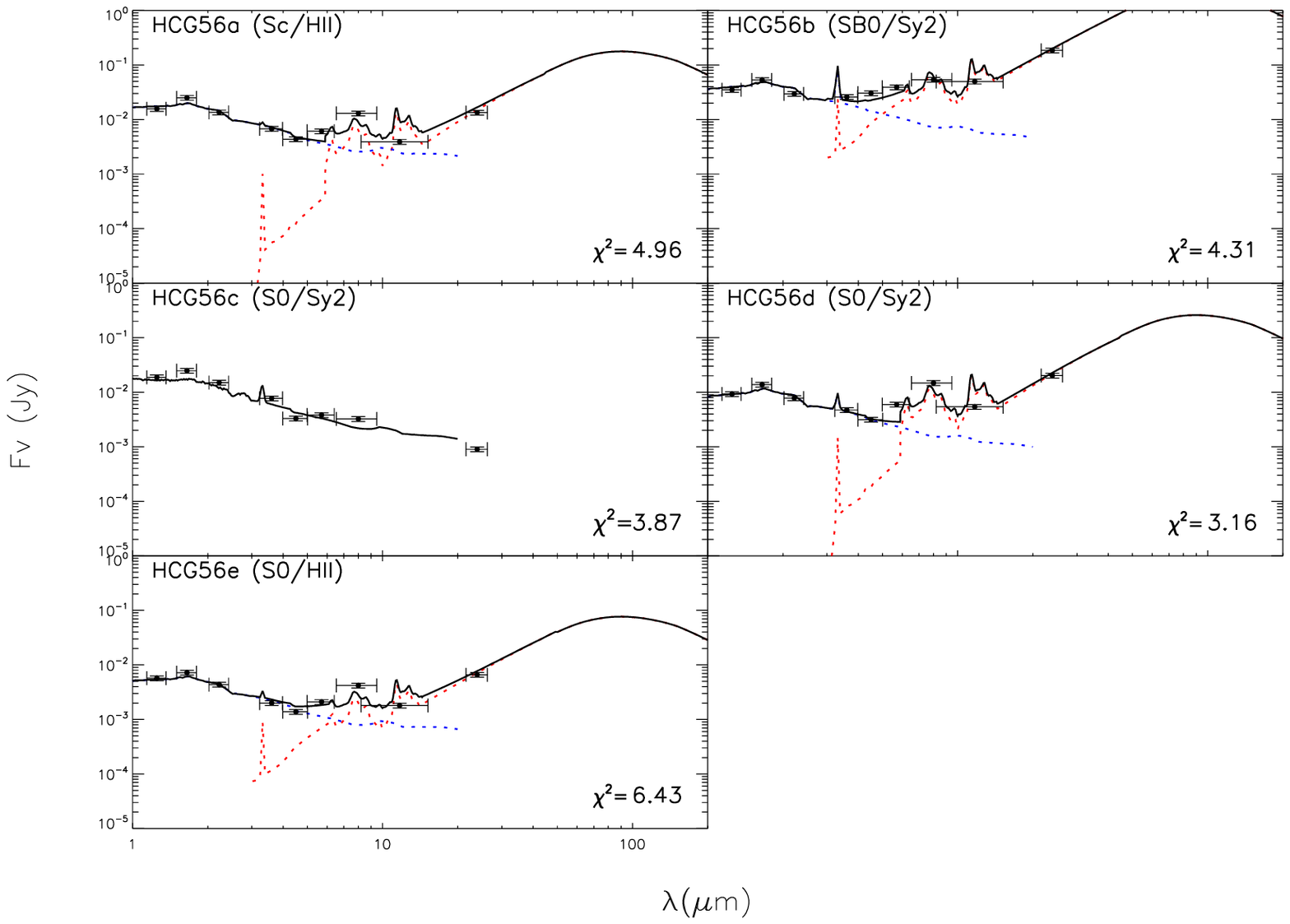}
\caption{Continued.}
\end{figure*}

\begin{figure*}
\setcounter{figure}{0} 
\includegraphics[scale=0.9]{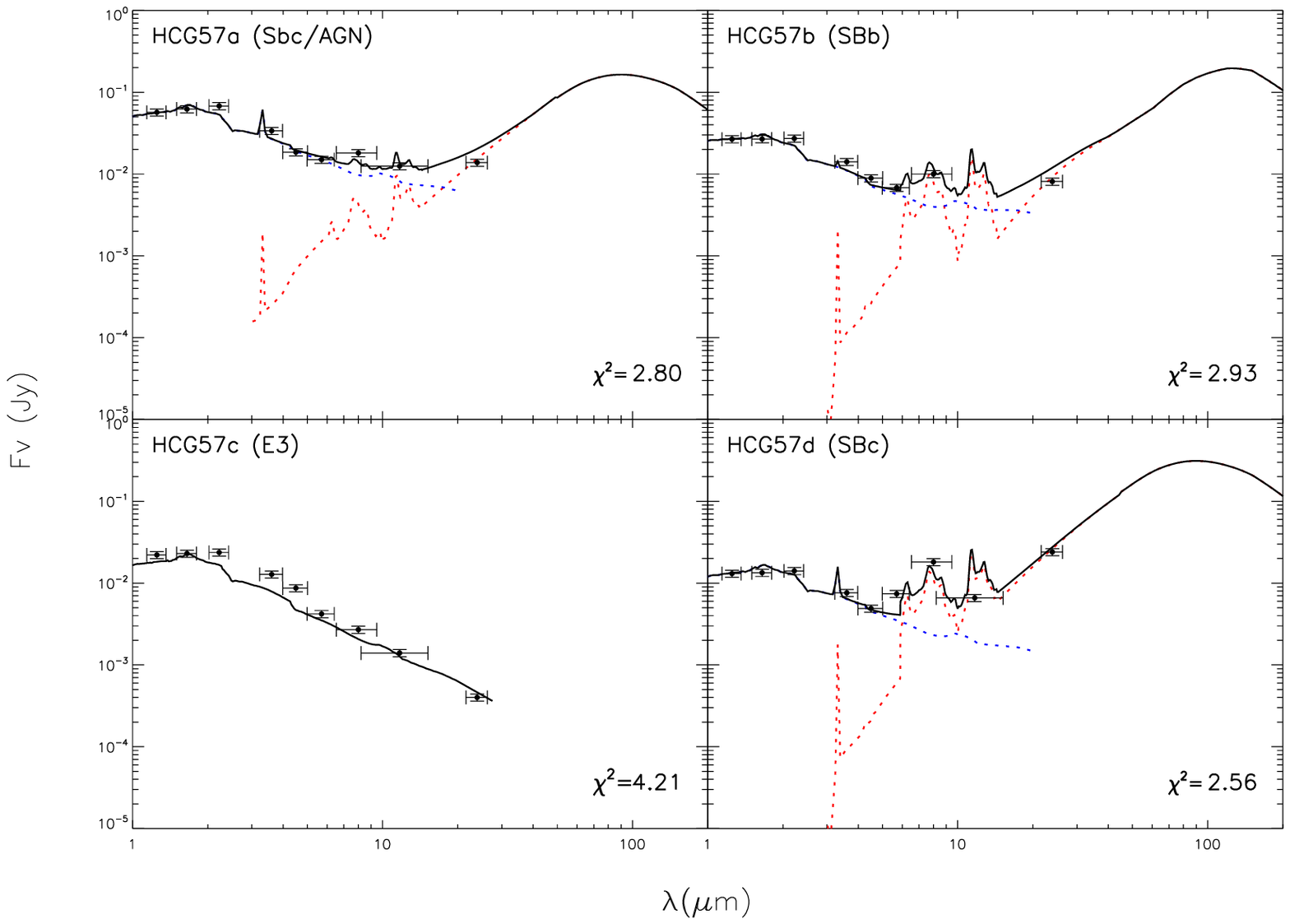}
\includegraphics[scale=0.9]{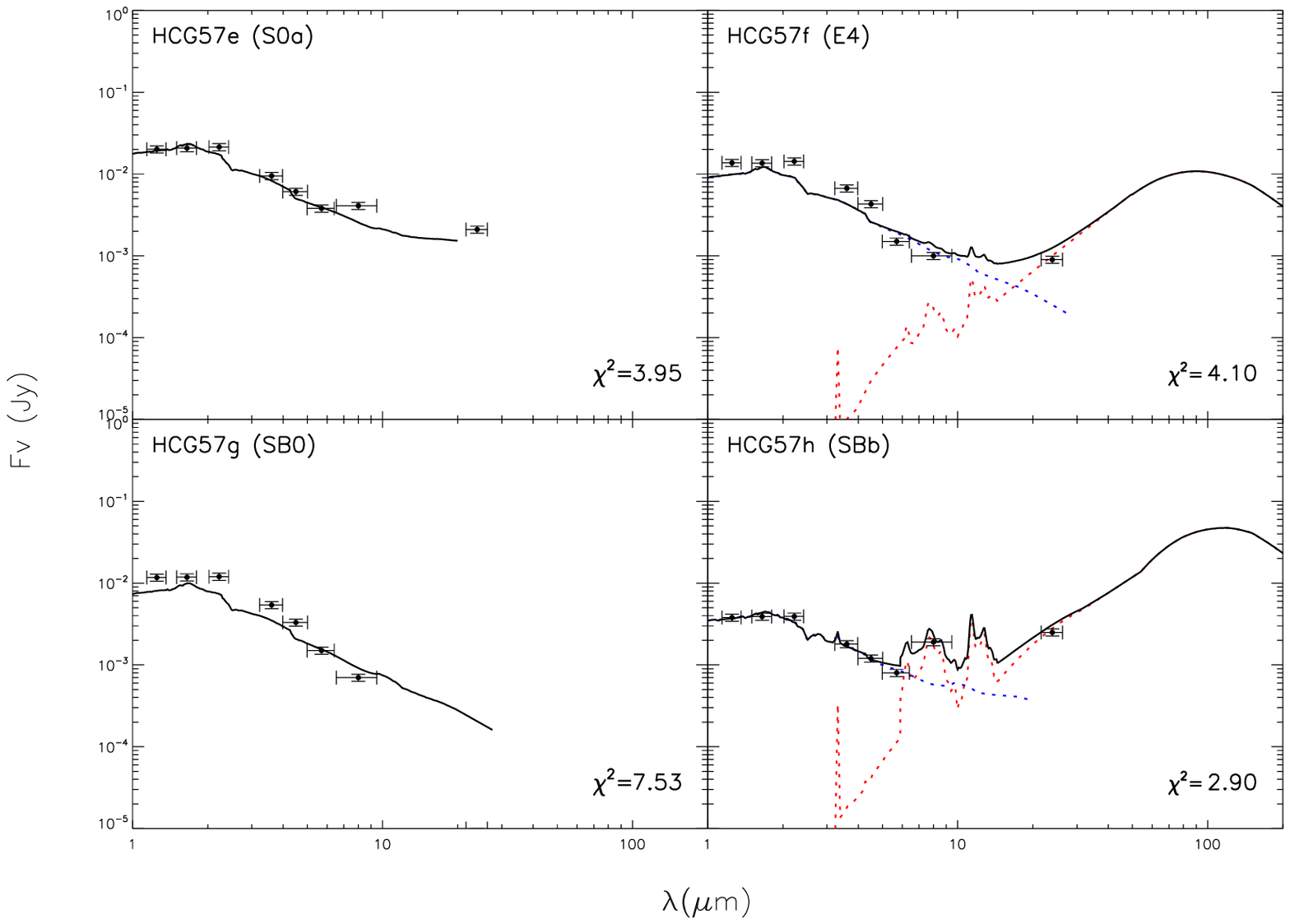}
\caption{Continued.}
\end{figure*}

\begin{figure*}
\setcounter{figure}{0} 
\includegraphics[scale=0.9]{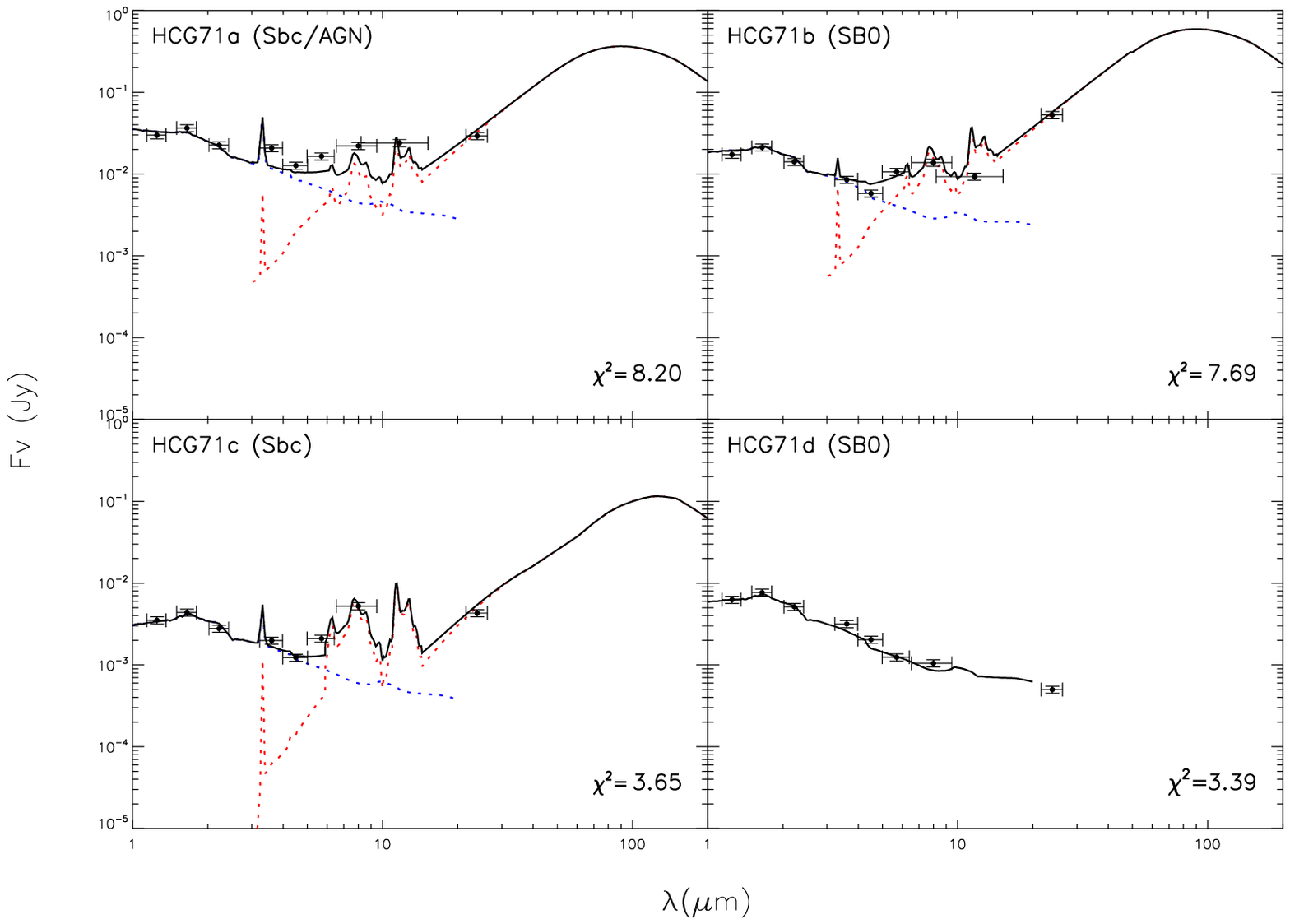}
\includegraphics[scale=0.9]{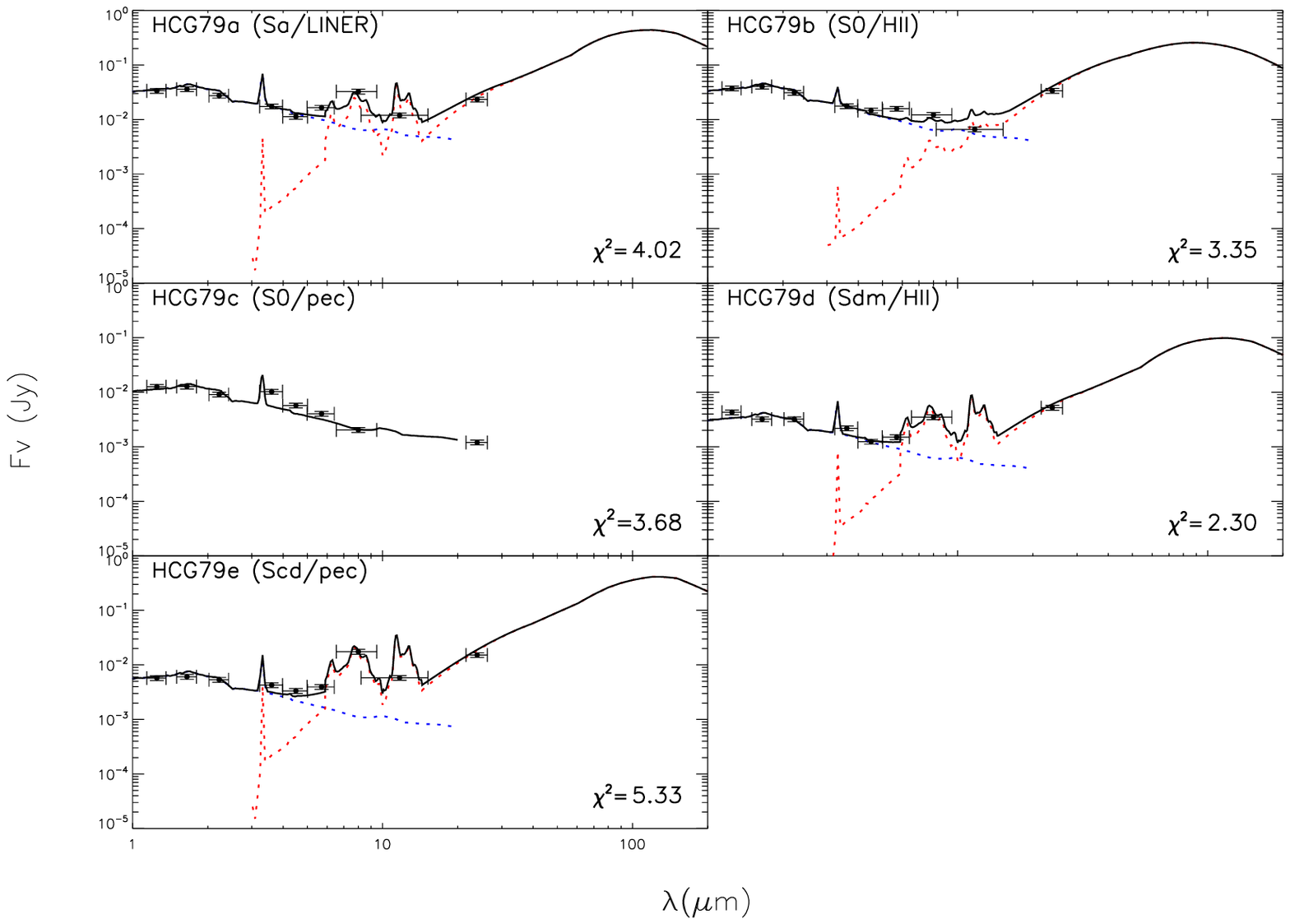}
\caption{Continued.}
\end{figure*}

\begin{figure*}
\setcounter{figure}{0} 
\includegraphics[scale=0.9]{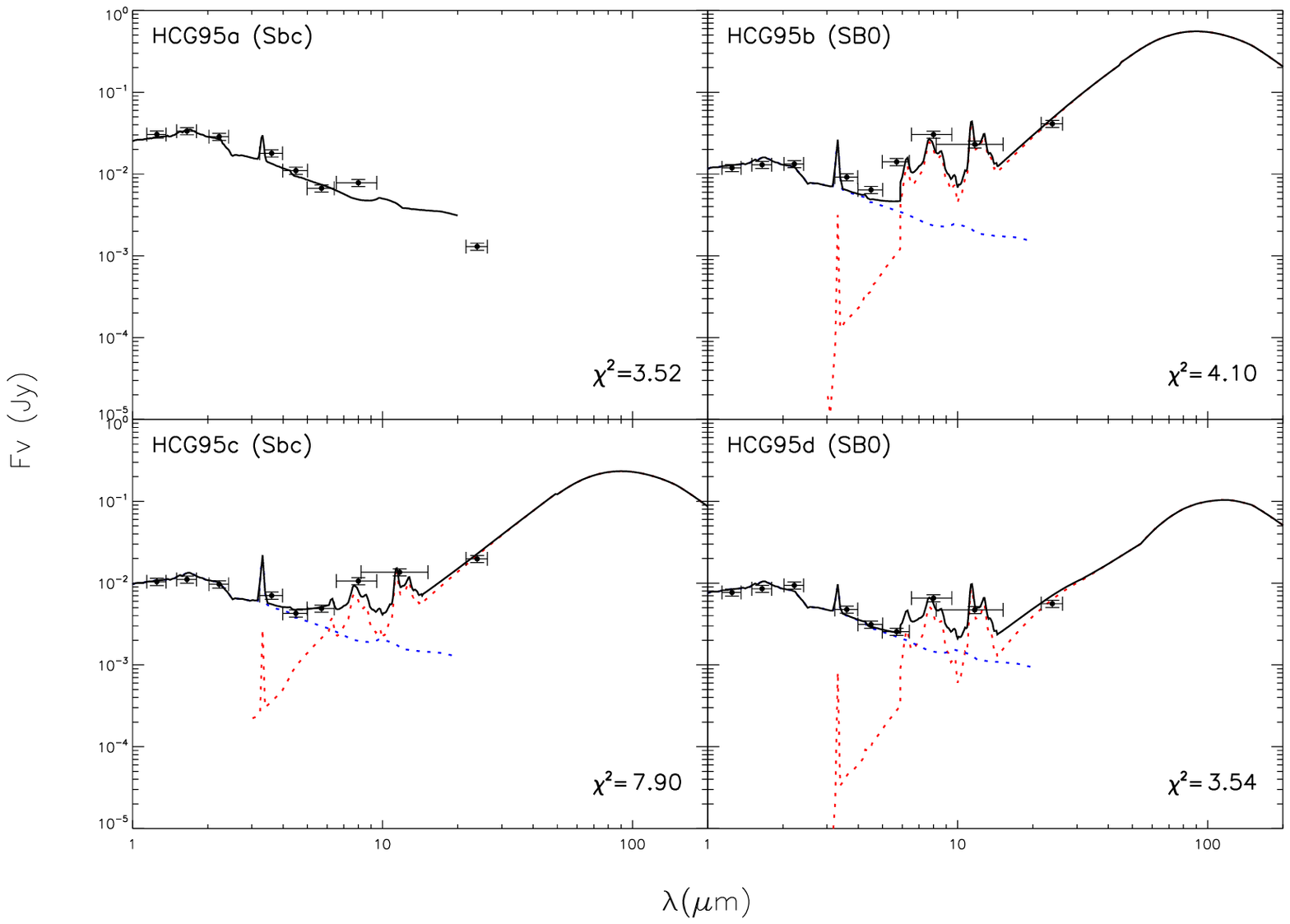}
\caption{Continued.}
\end{figure*}

\end{document}